\documentclass[aps,prd,reprint,a4paper,nofootinbib,amsmath,amsfonts,amssymb]{revtex4-1}
\usepackage[unicode=true,bookmarksopen=true,colorlinks,citecolor=blue,linkcolor=blue,urlcolor=blue]{hyperref}

\usepackage{mathtools}
\usepackage{lmodern}
\usepackage{dsfont}
\usepackage[T1]{fontenc}

\let\originalleft\left
\let\originalright\right
\renewcommand{\left}{\mathopen{}\mathclose\bgroup\originalleft}
\renewcommand{\right}{\aftergroup\egroup\originalright}

\newcommand{\mathe}{\mathrm{e}}
\newcommand{\mathi}{\mathrm{i}}
\newcommand{\total}{\mathop{}\!\mathrm{d}}
\newcommand{\abs}[1]{{\left\lvert{#1}\right\rvert}}
\newcommand{\eqend}[1]{\,#1}
\newcommand{\bigo}[1]{\mathcal{O}({#1})}
\newcommand{\bra}[1]{\left\langle{#1}\right\vert}
\newcommand{\ket}[1]{\left\vert{#1}\right\rangle}
\newcommand{\brst}{\mathop{}\!\mathsf{s}\hskip 0.05em\relax}
\newcommand{\sdot}{\!\cdot\!}

\usepackage{accents}
\newlength{\dhatheight}
\newcommand{\dhat}[1]{%
    \settoheight{\dhatheight}{\ensuremath{\hat{#1}}}%
    \addtolength{\dhatheight}{-0.25ex}%
    \hat{\vphantom{\rule{1pt}{\dhatheight}}%
    \smash{\hat{#1}}}}

\allowdisplaybreaks

\makeatletter
\def\@bibdataout@aps{%
 \immediate\write\@bibdataout{@CONTROL{apsrev41Control,author="08",editor="1",pages="1",title="0",year="0"}}%
 \if@filesw
  \immediate\write\@auxout{\string\citation{apsrev41Control}}%
 \fi
}%
\makeatother

\begin{document}

\title{Green's functions and Hadamard parametrices for vector and tensor fields in general linear covariant gauges}
\date{7. September 2017}

\author{Markus B. Fr{\"o}b}
\email{mbf503@york.ac.uk}
\affiliation{Department of Mathematics, University of York, Heslington, York, YO10 5DD, United Kingdom}

\author{Mojtaba \surname{Taslimi Tehrani}}
\email{mojtaba.taslimitehrani@mis.mpg.de}
\affiliation{Max-Planck-Institut f{\"u}r Mathematik in den Naturwissenschaften, Inselstr.~22, 04103 Leipzig, Germany}
\affiliation{Institut f{\"u}r Theoretische Physik, Universit{\"a}t Leipzig,  Br{\"u}derstr.~16, 04103 Leipzig, Germany}

\begin{abstract}
We determine the retarded and advanced Green's functions and Hadamard parametrices in curved spacetimes for linearized massive and massless gauge bosons and linearized Einstein gravity with a cosmological constant in general linear covariant gauges. These vector and tensor parametrices contain additional singular terms compared with their Feynman/de Donder-gauge counterpart. We also give explicit recursion relations for the Hadamard coefficients, and indicate their generalization to $n$ dimensions. Furthermore, we express the divergence and trace of the vector and tensor Green's functions in terms of derivatives of scalar and vector Green's functions, and show how these relations appear as Ward identities in the free quantum theory.
\end{abstract}

\maketitle

\section{Introduction}

A central notion in quantum field theory in a curved spacetime $M$ is that of Hadamard states~\cite{dewittbrehme1960,fullingsweenywald1978,kaywald1991}. These are a class of quantum states which exhibit physically reasonable properties, e.g., finite expectation values and fluctuations of the stress tensor, and they can be characterized by their singularity structure. For instance, the Minkowski vacuum, thermal states in Minkowski space and the Bunch--Davies states for free fields in cosmological spacetimes are Hadamard states, as well as all states which are obtained by applying smeared field operators to those states.

The short distance behavior of any Hadamard state is described by a Hadamard parametrix $H(x,x')$. This is a bi-solution of the corresponding field equation with a smooth source, which is defined locally and geometrically. That is, $H(x,x')$ is defined for all $x'$ in a convex normal neighborhood of $x$, and only depends on the geometry of $M$ in this neighborhood. In particular, it does not describe the two-point function of any particular preferred state, but instead specifies the singular part that any such state must have. Hadamard parametrices are used to define renormalized composite operators using the point-splitting method~\cite{christensen1976,christensen1978,wald1978}, which includes the renormalized stress tensor for scalars, spinors, vectors, gravitons and $p$-forms~\cite{brownottewill1986,allenfolacciottewill1988,folacci1991,moretti2001,decaninifolacci2008,hackmoretti2012,ambruswinstanley2015,kentwinstanley2015,belokognefolacci2016,belokognefolacciqueva2016}, in particular its trace anomaly~\cite{wald1978,hollandswald2005,dappiaggihackpinamonti2009}, and the calculation of chiral anomalies~\cite{qiuren1988,banerjeebanerjee1988,novotnyschnabl2000,zahn2015}. More generally, they play a crucial role in constructing the local and covariant time-ordered products on curved spacetimes~\cite{brunettifredenhagenkoehler1996,brunettifredenhagen2000,hollandswald2001,hollandswald2002}, which form the basis of renormalized perturbation theory on arbitrary (globally hyperbolic) curved backgrounds.

However, so far the treatment of theories with local gauge symmetry, namely Yang-Mills theories and linearized Einstein gravity, has been mostly restricted to special gauges: Feynman gauge in the case of Yang-Mills and de Donder gauge for linearized gravity. In these gauges, the equation of motion (EOM) is normally hyperbolic, i.e., the second derivatives only appear in form of the wave operator $\nabla^2 \equiv g^{\mu\nu} \nabla_\mu \nabla_\nu$. On a globally hyperbolic Lorentzian manifold, normally hyperbolic operators have a well-posed Cauchy problem (see, e.g., Ref.~\cite{baerginouxpfaeffle2007}), and consequently there exist unique retarded and advanced Green's functions and corresponding Hadamard parametrices. In more general gauges, the differential operators appearing in the EOM contain second derivatives other than $\nabla^2$, and are only Green hyperbolic~\cite{baerginoux2012,baer2015}. For Green hyperbolic operators, while uniqueness of Green's functions still holds, their existence is not guaranteed.

The main aim of the present article is to construct explicit Green's functions and the corresponding Hadamard parametrices in curved spacetimes for vector gauge bosons and linearized Einstein gravity with a cosmological constant in general linear covariant gauges. For the vector theory, there is a family of linear covariant gauges with parameter $\xi \in \mathbb{R}$\footnote{While the limit $\xi \to 0$ is not defined for the differential operator, it exists for the Green's function~\eqref{vector_propagator}, as is well known.}, in which the differential operator appearing in the EOM reads
\begin{equation}
\label{vector_p_def}
P_{m^2,\xi}^{\mu\nu} \equiv g^{\mu\nu} \left( \nabla^2 - m^2 \right) - R^{\mu\nu} - \frac{\xi-1}{\xi} \nabla^\mu \nabla^\nu \eqend{,}
\end{equation}
where we have included a mass term $m^2$ arising from spontaneous symmetry breaking. Because of the last term, this operator is not normally hyperbolic for general $\xi$, but only for $\xi = 1$ (the Feynman gauge). For linearized gravity, there is a two-parameter family of linear covariant gauges parametrized by $\xi, \zeta \in \mathbb{R}$ with the corresponding operator $P_{\xi,\zeta}^{\mu\nu\rho\sigma}$ given in Eq.~\eqref{tensor_p_def}. Again, this operator is not normally hyperbolic in general, and only in the gauge $\xi = \zeta = 1$ (the de Donder gauge) the existence of Green's functions is guaranteed\footnote{Although not even $P_{1,1}^{\mu\nu\rho\sigma}$ is normally hyperbolic, its trace-reversed version $\bar{P}_{1,1}^{\mu\nu\rho\sigma} \equiv P_{1,1}^{\mu\nu\rho\sigma} - 1/(n-2) g^{\rho\sigma} g_{\alpha\beta} P_{1,1}^{\mu\nu\alpha\beta}$ is, and one can reconstruct the Green's functions of $P_{1,1}^{\mu\nu\rho\sigma}$ from the ones of $\bar{P}_{1,1}^{\mu\nu\rho\sigma}$ by purely algebraic means, see Eq.~\eqref{tensor_green_tracereversed}.}.

For practical calculations it is important to know the vector and tensor Green's functions and Hadamard parametrices with arbitrary $\xi$ and $\zeta$, since the calculations may considerably simplify in certain gauges. For example, in Landau gauge $\xi \to 0$ (and $\zeta \to \infty$ for the tensor case) the divergence of the Green's function or Feynman propagator vanishes, and the gauge $\xi \to 0$, $\zeta \to n/2$ (where $n$ is the dimension of $M$) presents advantages in AdS/CFT calculations~\cite{dhokeretal1999}. Moreover, keeping the gauge parameters arbitrary serves as a consistency check of gauge-fixing independence in practical calculations, since all terms depending on $\xi$ or $\zeta$ must cancel out in the final results for physical quantities.

Let us explain our general strategy by considering the massive Proca operator~\cite{proca1936}
\begin{equation}
\label{vector_proca}
P_{m^2,\infty}^{\mu\nu} = g^{\mu\nu} \left( \nabla^2 - m^2 \right) - R^{\mu\nu} - \nabla^\mu \nabla^\nu \eqend{,}
\end{equation}
which is a prototype of a Green hyperbolic operator, obtainable as the limit $\xi \to \infty$ of Eq.~\eqref{vector_p_def}. It is known~\cite{baerginoux2012} that the (advanced or retarded) Green's function $G^{m^2,\infty}_{\mu\rho'}(x,x')$ of the Proca operator can be constructed from the massive vector Green's function in Feynman gauge $G^{m^2,1}_{\mu\rho'}(x,x')$ (which is known to exist and be unique since $P_{m^2,1}^{\mu\nu}$ is normally hyperbolic) and the Green's function $G_{m^2}(x,x')$ of the massive Klein--Gordon operator for scalar fields according to
\begin{equation}
\label{proca_green}
G^{m^2,\infty}_{\mu\rho'}(x,x') = G^{m^2,1}_{\mu\rho'}(x,x') + \frac{1}{m^2} \nabla_\mu \nabla_{\rho'} G_{m^2}(x,x') \eqend{.}
\end{equation}
Similarly, we can express the Green's function of $P_{m^2,\xi}^{\mu\nu}$ in terms of the Green's functions $G^{m^2,1}_{\mu\rho'}$, $G_{m^2}$ and $G_{\xi m^2}$. It turns out that, contrary to Eq.~\eqref{proca_green}, $G^{m^2,\xi}_{\mu\rho'}$ for $\xi < \infty$ admits a well-defined massless limit, and $G^{0,\xi}_{\mu\rho'}$ can be expressed using mass derivatives of $G_{m^2}$ at $m = 0$.

Once we have obtained an expression for the advanced or retarded Green's function, we construct the corresponding Hadamard parametrix. Let us start with the scalar case, and assume given a Hadamard state $\ket{\psi}$. Then the Wightman function\footnote{The prefactor for the Wightman function, the Feynman propagator and various other Green's functions is a matter of convention.}
\begin{equation}
\label{scalar_wightman}
G^+_{m^2}(x,x') \equiv - \mathi \bra{\psi} \phi(x) \phi(x') \ket{\psi}
\end{equation}
is a solution to $P_{m^2} G^+_{m^2}(x,x') = 0$ with the Klein--Gordon operator
\begin{equation}
\label{scalar_p_def}
P_{m^2} \equiv \nabla^2 - m^2 \eqend{,}
\end{equation}
while the time-ordered Feynman propagator
\begin{equation}\begin{split}
\label{scalar_feynman}
G^\text{F}_{m^2}(x,x') &\equiv - \mathi \bra{\psi} \mathcal{T} \phi(x) \phi(x') \ket{\psi} \\
&= \Theta(t-t') G^+_{m^2}(x,x') + \Theta(t'-t) G^+_{m^2}(x',x)
\end{split}\end{equation}
is a solution to the inhomogeneous equation $P_{m^2} G^\text{F}_{m^2}(x,x') = \delta(x,x')$ with the covariant $\delta$ distribution $\delta(x,x') \equiv \delta^4(x-x')/\sqrt{-g}$, and where $t = t(x)$ is any time function. From those, the retarded Green's function can be obtained via
\begin{equation}\begin{split}
\label{scalar_retarded}
G^\text{ret}_{m^2}(x,x') &\equiv \Theta(t-t') \left[ G^+_{m^2}(x,x') - G^+_{m^2}(x',x) \right] \\
&= G^\text{F}_{m^2}(x,x') - G^+_{m^2}(x',x) \eqend{.}
\end{split}\end{equation}
Since $\ket{\psi}$ is a Hadamard state, the Wightman function~\eqref{scalar_wightman} in four dimensions locally takes the form~\cite{hadamard1932,baerginouxpfaeffle2007}
\begin{widetext}
\begin{equation}
\label{wightman_scalar}
G^+_{m^2}(x,x') = - \frac{\mathi}{8 \pi^2} \left[ \frac{U_{m^2}(x,x')}{\sigma(x,x') + \mathi \epsilon (t-t')} + V_{m^2}(x,x') \ln\left[ \mu^2 \sigma(x,x') + \mathi \epsilon (t-t') \right] + W_{m^2}(x,x') \right] \eqend{,}
\end{equation}
\end{widetext}
where $\mu$ is a mass scale, the functions $U_{m^2}$, $V_{m^2}$ and $W_{m^2}$ are smooth symmetric biscalars, and the distributional limit $\epsilon \to 0^+$ is understood. The symmetric biscalar $\sigma(x,x')$ is Synge's world function~\cite{synge1960}, equal to one half of the (signed) square of the geodesic distance between $x$ and $x'$, which is well defined locally (i.e., when $x'$ is in a normal geodesic neighborhood of $x$). It is easy to check that the Feynman propagator~\eqref{scalar_feynman} is given by expression~\eqref{wightman_scalar} with $\sigma(x,x') + \mathi \epsilon (t-t')$ replaced by $\sigma(x,x') + \mathi \epsilon$. The retarded Green's function~\eqref{scalar_retarded} thus reads
\begin{equation}
\label{hadamard_4d_retarded}
G^\text{ret}_{m^2}(x,x') = - \frac{\Theta(t-t')}{4 \pi} \left[ U_{m^2} \delta(\sigma) - V_{m^2} \Theta(-\sigma) \right] \eqend{,}
\end{equation}
using the well-known formulas (valid in the distributional limit $\epsilon \to 0^+$)
\begin{subequations}\label{sokhotski_plemelj}\begin{align}
\frac{1}{x + \mathi \epsilon} &= \mathcal{P}\!f \frac{1}{x} - \mathi \pi \delta(x) \eqend{,} \\
\ln (x + \mathi \epsilon) &= \ln \abs{x} + \mathi \pi \Theta(-x) \eqend{,}
\end{align}\end{subequations}
with Hadamard's finite part distribution $\mathcal{P}\!f$. Since the retarded Green's function is unique, it doesn't depend on the state $\ket{\psi}$, and thus in particular the biscalars $U_{m^2}$ and $V_{m^2}$ are state-independent and only depend on the geometry of the spacetime $M$. Therefore, we see that the biscalar $W_{m^2}$ encodes the dependence of the Wightman function $G^+_{m^2}$~\eqref{wightman_scalar} or the Feynman propagator $G^\text{F}_{m^2}$~\eqref{scalar_feynman} on the quantum state $\ket{\psi}$. On the other hand, the geometric state-independent Hadamard parametrix $H_{m^2}(x,x')$ is given by
\begin{equation}
\label{hadamard_4d_scalar}
H_{m^2} \equiv - \frac{\mathi}{8 \pi^2} \left[ \frac{U_{m^2}}{\sigma_\epsilon} + V_{m^2} \ln\left( \mu^2 \sigma_\epsilon \right) \right] \eqend{,}
\end{equation}
where the proper $\epsilon$ prescription depends on whether one considers the Wightman function $G^+_{m^2}$ or the Feynman propagator $G^\text{F}_{m^2}$. In the Wightman case, the Hadamard parametrix $H^+_{m^2}(x,x')$ has the same $\epsilon$ prescription as the Wightman function~\eqref{wightman_scalar}, namely it is of the form~\eqref{hadamard_4d_scalar} with $\sigma_\epsilon = \sigma + \mathi \epsilon (t-t')$, while the Feynman Hadamard parametrix $H^\text{F}_{m^2}(x,x')$ is of the form~\eqref{hadamard_4d_scalar} with the Feynman prescription $\sigma_\epsilon = \sigma + \mathi \epsilon$. For notational convenience, in the following we will work with the generic form~\eqref{hadamard_4d_scalar}, and moreover drop the $\epsilon$ subscript from $\sigma_\epsilon$. Note that both the Wightman function $G^+_{m^2}$ and the Feynman propagator $G^\text{F}_{m^2}$ are actually independent of the scale $\mu$, and that a change $\mu \to \mu'$ is compensated by the change $W_{m^2}(x,x') \to W_{m^2}(x,x') + 2 V_{m^2}(x,x') \ln\left( \mu/\mu' \right)$. For massive scalars, one could choose $\mu$ to be equal to $m$, which however creates problems as $m \to 0$ for the Hadamard parametrix.

In the vector case in Feynman gauge $\xi = 1$, the Hadamard parametrix is similarly given by~\cite{dewittbrehme1960}
\begin{equation}
\label{hadamard_parametrix_vector_feynman}
H^{m^2,1}_{\mu\rho'}(x,x') \equiv - \frac{\mathi}{8 \pi^2} \left[ \frac{U^{m^2,1}_{\mu\rho'}}{\sigma} + V^{m^2,1}_{\mu\rho'} \ln\left( \mu^2 \sigma \right) \right] \eqend{,}
\end{equation}
where the same remarks about the proper $\epsilon$ prescription apply. The Wightman function now contains a state-dependent bitensor $W^{m^2,1}_{\mu\rho'}$. As we will see [cf. Eq.~\eqref{hadamard_4d_vector_xi}], for a general gauge $\xi \neq 1$ the Hadamard parametrix contains additional singular terms proportional to $\sigma^{-2}$, and a new set of bitensors $\{ U, V \}^{m^2,\xi}_{\mu\rho'}$ related to the Feynman gauge bitensors and the biscalars appearing in the scalar Hadamard parametrix~\eqref{hadamard_4d_scalar}.

Along the same lines, we calculate the retarded or advanced Green's function $G^{\xi,\zeta}_{\mu\nu\rho'\sigma'}$ for linearized gravity around backgrounds which are solutions to Einstein's equations with a cosmological constant $\Lambda$. We express $G^{\xi,\zeta}_{\mu\nu\rho'\sigma'}$ in terms of the de Donder gauge Green's function $G^{1,1}_{\mu\nu\rho'\sigma'}$, and the vector and scalar Green's functions $G^{\mathfrak{m}^2,1}_{\mu\rho'}$ and $G_{\mathfrak{m}^2}$ with $\mathfrak{m}$ a mass parameter proportional to $\Lambda$, and we show that the limit $\mathfrak{m}^2 \rightarrow 0$ (which corresponds to $\Lambda = 0$) exists. While for de Donder gauge the most singular term in the Hadamard parametrix is again proportional to $\sigma^{-1}$, for a general gauge it turns out to be proportional to $\sigma^{-3}$.

The remainder of the article is organized as follows: in Sec.~\ref{section_green} we determine the advanced and retarded Green's functions for vector and tensor fields in general linear and covariant gauges, in Sec.~\ref{section_ward} we show that certain divergence and trace identities which follow for the Green's functions can be obtained as Ward identities in the free quantum theory, and in Sec.~\ref{section_hadamard} we determine the corresponding Hadamard parametrices. We close in Sec.~\ref{section_outlook} with an outlook on future work, and, with a view on practical applications, also give expressions for the state-dependent $W$ coefficients in Appendix~\ref{appendix_w} under the assumption that the Wightman functions or Feynman propagators in general linear and covariant gauges are determined in the same way as the Green's functions. We use the ``+++'' sign convention of Ref.~\cite{mtw}, and work in $n \geq 2$ dimensions, except for the Hadamard expansion in Sec.~\ref{section_hadamard} where we restrict to four dimensions.

\section{Green's functions}
\label{section_green}

Given a differential operator $P$, by definition advanced and retarded Green's functions $G^\text{adv/ret}$ satisfy
\begin{equation}
P_x G^\text{adv/ret}(x,x') = \delta(x,x') = P_{x'} G^\text{adv/ret}(x,x')
\end{equation}
with the support properties
\begin{equation}\begin{split}
\operatorname{supp} \int G^\text{ret}(x,x') f(x') \total_g x' \subset J^+(\operatorname{supp} f) \eqend{,} \\
\operatorname{supp} \int G^\text{adv}(x,x') f(x') \total_g x' \subset J^-(\operatorname{supp} f)
\end{split}\end{equation}
for any compactly supported test function $f$, where $J^+(S)$ [$J^-(S)$] is the causal future (past) of a set $S$, and we set $\total_g x \equiv \sqrt{-g} \total^n x$. Given a Green's function, the solution of the inhomogeneous equation $P \phi = f$ with retarded or advanced boundary conditions is then given by the formula
\begin{equation}
\phi^\text{ret/adv}(x) = \int G^\text{ret/adv}(x,x') f(x') \total_g x' \eqend{.}
\end{equation}
Uniqueness of Green's functions with a specific boundary condition, thus, leads to unique solutions to the inhomogeneous equation under those boundary conditions. In particular, for vanishing source $f = 0$ we obtain $\phi = 0$, which we will make use of in the following.

In this section, we determine the advanced and retarded Green's functions of vector and tensor fields in general linear covariant gauges. Since the Green’s functions that we obtain are expressed in terms of the scalar and vector Green's function and their mass derivatives, we begin with the massive Klein--Gordon operator. Our calculations are valid for either advanced or retarded boundary conditions, and for ease of notation we drop the superscript ``adv/ret'' in the remainder of the article.

\subsection{Scalar field}

Consider a massive scalar field with minimal curvature coupling. The differential operator appearing in its EOM is the Klein--Gordon operator $P_{m^2}$ defined in Eq.~\eqref{scalar_p_def}. As discussed in the introduction, $P_{m^2}$ is a normally hyperbolic differential operator, and thus admits unique advanced and retarded Green's functions $G_{m^2}(x,x')$ which satisfy
\begin{equation}
\label{scalar_kleingordon}
P_{m^2} G_{m^2}(x,x') = \delta(x,x') \eqend{,}
\end{equation}
with the support properties discussed above. For later use, we will need mass derivatives of Green's functions, and we define
\begin{equation}
\label{scalar_massderivative}
\hat{G}_{M^2} \equiv \frac{\partial G_{m^2}}{\partial m^2} \bigg\rvert_{m^2 = M^2} \eqend{,} \qquad \dhat{G}_{M^2} \equiv \frac{\partial \hat{G}_{m^2}}{\partial m^2} \bigg\rvert_{m^2 = M^2} \eqend{.}
\end{equation}
By differentiating Eq.~\eqref{scalar_kleingordon} with respect to the mass, we obtain
\begin{equation}
\label{scalar_kleingordon_massderiv}
P_{m^2} \hat{G}_{m^2}(x,x') = G_{m^2}(x,x')
\end{equation}
and
\begin{equation}
\label{scalar_kleingordon_massderiv2}
P_{m^2} \dhat{G}_{m^2}(x,x') = 2 \hat{G}_{m^2}(x,x') \eqend{.}
\end{equation}

\subsection{Vector field}

As explained in the introduction, the linearized Yang-Mills equation is not normally hyperbolic. Here, we consider a (massive) vector field in a general linear covariant gauge, whose differential operator $P_{m^2,\xi}^{\mu\nu}$ is given by Eq.~\eqref{vector_p_def}, with $\xi$ a gauge parameter. The choice $\xi = 1$ (Feynman gauge) eliminates the last term in Eq.~\eqref{vector_p_def} and $P_{m^2,1}^{\mu\nu}$ is a normally hyperbolic operator. The Green's functions that we want to determine satisfy
\begin{equation}
\label{vector_eom}
P_{m^2,\xi}^{\mu\nu} G^{m^2,\xi}_{\nu\beta'}(x,x') = g^\mu_{\beta'} \delta(x,x') \eqend{,}
\end{equation}
with the bitensor of parallel transport $g_{\mu\beta'}$, which for coinciding points reduces to the metric
\begin{equation}
\lim_{x' \to x} g_{\mu\beta'} = g_{\mu\beta} \eqend{.}
\end{equation}

\subsubsection*{Divergence identity for \texorpdfstring{$\xi=1$}{xi=1}}

To obtain the Green's function $G^{m^2,\xi}_{\nu\beta'}(x,x')$, we first need to determine an expression for the divergence of the Feynman gauge Green's function $G^{m^2,1}_{\nu\beta'}(x,x')$. We follow Ref.~\cite{dewittbrehme1960} and calculate
\begin{equation}\begin{split}
&P_{m^2} \left[ \nabla^\nu G^{m^2,1}_{\nu\beta'}(x,x') + \nabla_{\beta'} G_{m^2}(x,x') \right] \\
&\quad= \nabla_\mu \left[ P_{m^2,1}^{\mu\nu} G^{m^2,1}_{\nu\beta'}(x,x') \right] + \nabla_{\beta'} P_{m^2} G_{m^2}(x,x') \\
&\quad= \nabla_\mu \left[ g^\mu_{\beta'} \delta(x,x') \right] + \nabla_{\beta'} \delta(x,x') = 0 \eqend{,}
\end{split}\end{equation}
where the equality of the last line follows from the properties of the parallel propagator $g_{\mu\beta'}$~\cite{synge1960,dewittbrehme1960}.

According to the previous discussion, since $P_{m^2}$ is normally hyperbolic the solution of this equation with either retarded or advanced boundary conditions is unique and thus vanishing, and we infer that
\begin{equation}
\label{vector_feyn_div}
\nabla^\nu G^{m^2,1}_{\nu\beta'}(x,x') = - \nabla_{\beta'} G_{m^2}(x,x') \eqend{.}
\end{equation}
This relation is in fact necessary for the Ward identities to hold in the (free) quantum theory, which we explain in the framework of BRST quantization in Sec.~\ref{section_ward}.

\subsubsection*{The massive vector Green's function for general \texorpdfstring{$\xi$}{xi}}

In the general case, inspired from the known flat-space Green's function we consider the combination
\begin{equation}
\tilde{G}^{m^2,\xi}_{\nu\beta'} \equiv G^{m^2,\xi}_{\nu\beta'} + (\xi-1) \nabla_\nu \nabla_{\beta'} \tilde{G}
\end{equation}
with an unspecified function $\tilde{G}$. A short calculation shows that
\begin{equation}
\label{vector_xi_p1_tilde_eq1}
P_{m^2,1}^{\mu\nu} \tilde{G}^{m^2,\xi}_{\nu\beta'}(x,x') = g^\mu_{\beta'} \delta(x,x')
\end{equation}
if $\tilde{G}$ fulfils
\begin{equation}
\label{vector_xi_p1_tilde_eq2}
\nabla^\mu \left[ \nabla_{\beta'} P_{\xi m^2} \tilde{G}(x,x') + \nabla^\nu \tilde{G}^{m^2,\xi}_{\nu\beta'}(x,x') \right] = 0 \eqend{.}
\end{equation}
Since $P_{m^2,1}^{\mu\nu}$ is normally hyperbolic, the solution of Eq.~\eqref{vector_xi_p1_tilde_eq1} with retarded or advanced boundary conditions is unique, and we conclude that $\tilde{G}^{m^2,\xi}_{\nu\beta'} = G^{m^2,1}_{\nu\beta'}$. The relation~\eqref{vector_feyn_div} then shows that Eq.~\eqref{vector_xi_p1_tilde_eq2} for $\tilde{G}$ reduces to
\begin{equation}
\label{vector_xi_scalar_eq}
\nabla^\mu \nabla_{\beta'} \left[ P_{\xi m^2} \tilde{G}(x,x') - G_{m^2}(x,x') \right] = 0 \eqend{.}
\end{equation}
One easily verifies that a solution of this equation is given by
\begin{equation}
\tilde{G} = \frac{G_{m^2} - G_{\xi m^2}}{(1-\xi) m^2} \eqend{,}
\end{equation}
and we thus obtain
\begin{equation}
\label{vector_propagator}
G^{m^2,\xi}_{\nu\beta'} = G^{m^2,1}_{\nu\beta'} + \frac{1}{m^2} \nabla_\nu \nabla_{\beta'} \left( G_{m^2} - G_{\xi m^2} \right) \eqend{.}
\end{equation}
For the divergence of the general vector Green's function, we then calculate using the relation~\eqref{vector_feyn_div}
\begin{equation}
\label{vector_xi_div}
\nabla^\nu G^{m^2,\xi}_{\nu\beta'} = - \xi \nabla_{\beta'} G_{\xi m^2} \eqend{,}
\end{equation}
which reduces to relation~\eqref{vector_feyn_div} for $\xi = 1$. We see that the transversality of the Green's function in Landau gauge $\xi = 0$, which is known from the flat-space case, holds also in general spacetimes, i.e.,
\begin{equation}
\nabla^\nu G^{m^2,0}_{\nu\beta'} = 0 \eqend{.}
\end{equation}

\subsubsection*{The massless limit}

In the limit $m \to 0$, we have
\begin{equation}
G_{m^2} = G_0 + m^2 \hat{G}_0 + \bigo{m^4} \eqend{,}
\end{equation}
and we obtain
\begin{equation}
\label{vector_propagator_massless}
G^{0,\xi}_{\nu\beta'} = G^{0,1}_{\nu\beta'} - (\xi-1) \nabla_\nu \nabla_{\beta'} \hat{G}_0 \eqend{,}
\end{equation}
which also can be obtained from the direct solution of Eq.~\eqref{vector_xi_scalar_eq} using Eq.~\eqref{scalar_kleingordon_massderiv} for $m^2 = 0$.

\subsection{Tensor field}

Lastly we would like to determine the Green's functions for a symmetric second-rank tensor field (which we call graviton) subject to the linearized Einstein equation, which is the basic quantum field in perturbative quantum gravity around fixed backgrounds~\cite{minosasakitanaka1997,fewsterhunt2013,brunettifredenhagenreijzner2016}. Since in two dimensions the integral of the Ricci scalar is a topological invariant, we restrict to $n \geq 3$ dimensions.
In this work, we consider background metrics $g_{\mu\nu}$ which satisfy Einstein's equation with a cosmological constant\footnote{The generalization to other background fields, such as the inflaton in cosmological spacetimes, proceeds along the same lines.}
\begin{equation}
\label{einstein_eq}
R_{\mu\nu} - \frac{1}{2} g_{\mu\nu} R + g_{\mu\nu} \Lambda = 0 \eqend{,}
\end{equation}
which implies
\begin{equation}
R_{\mu\nu} = \frac{2 \Lambda}{n-2} g_{\mu\nu} \eqend{,} \qquad R = \frac{2 n \Lambda}{n-2} \eqend{,} \qquad \nabla^\mu R_{\mu\nu\rho\sigma} = 0 \eqend{,}
\end{equation}
where the last equation follows from the Bianchi identities. We stress that apart from this conditions, the Riemann tensor is unconstrained; in particular we are \emph{not} restricting to de~Sitter or anti-de~Sitter spacetime where the Weyl tensor $C_{\mu\nu\rho\sigma} = R_{\mu\nu\rho\sigma} - 2/(n-2) \left( R_{\mu[\rho} g_{\sigma]\nu} - R_{\nu[\rho} g_{\sigma]\mu} \right) + 2/[(n-2)(n-1)] R g_{\mu[\rho} g_{\sigma]\nu}$ would vanish.

We do not consider a mass term for the graviton. Writing the perturbed metric as background $g_{\mu\nu}$ plus perturbation $\kappa h_{\mu\nu}$ with $\kappa^2 = 16 \pi G_\text{N}$, and expanding the Einstein-Hilbert action with cosmological constant to second order in $h_{\mu\nu}$, we obtain the action
\begin{equation}\begin{split}
\label{graviton_action}
S^{(0)} = \frac{1}{4} \int &\bigg[ h^{\mu\nu} \left( \nabla^2 h_{\mu\nu} - 2 \nabla^\rho \nabla_\mu h_{\nu\rho} + 2 \nabla_\mu \nabla_\nu h \right) \\
&\quad- h \nabla^2 h + \frac{2}{n-2} \left( 2 h^{\mu\nu} h_{\mu\nu} - h^2 \right) \Lambda \bigg] \total_g x \eqend{,}
\end{split}\end{equation}
where $h \equiv g^{\mu\nu} h_{\mu\nu}$ and $\nabla_\mu$ is the covariant derivative operator with respect to the background metric $g_{\mu\nu}$. To perform gauge fixing, we add to $S^{(0)}$ the action
\begin{equation}
\label{gauge-fixing-action}
S_\text{GF} = - \frac{1}{2 \xi} \int \left( \nabla^\nu h_{\mu\nu} - \frac{\nabla_\mu h}{2 \zeta} \right) \left( \nabla_\rho h^{\mu\rho} - \frac{\nabla^\mu h}{2 \zeta} \right) \total_g x
\end{equation}
with two gauge parameters $\xi$ and $\zeta$, where the analogue of Feynman gauge, usually called de Donder gauge, is given by $\xi = \zeta = 1$. Variations of $S^{(0)} + S_{\text{GF}}$ with respect to $h_{\mu\nu}$ leads to the field equation $P_{\xi,\zeta}^{\rho\sigma\mu\nu} h_{\mu\nu} = 0$, with the differential operator
\begin{widetext}
\begin{equation}\begin{split}
\label{tensor_p_def}
P_{\xi,\zeta}^{\rho\sigma\mu\nu} &\equiv \frac{1}{2} \left( g^{\rho(\mu} g^{\nu)\sigma} - \frac{1}{2} g^{\rho\sigma} g^{\mu\nu} \right) \nabla^2 + R^{\rho\gamma\sigma\delta} \left( \delta_\gamma^{(\mu} \delta^{\nu)}_\delta - \frac{1}{2} g_{\gamma\delta} g^{\mu\nu} \right) - \left( 1 - \frac{1}{\xi} \right) \nabla^{(\rho} g^{\sigma)(\mu} \nabla^{\nu)} \\
&\quad+ \frac{1}{2} \left( 1 - \frac{1}{\xi \zeta} \right) \left( g^{\mu\nu} \nabla^\rho \nabla^\sigma + g^{\rho\sigma} \nabla^\mu \nabla^\nu \right) - \frac{1}{4} \left( 1 - \frac{1}{\xi \zeta^2} \right) g^{\rho\sigma} g^{\mu\nu} \nabla^2 \eqend{,}
\end{split}\end{equation}
\end{widetext}
and the Green's function $G^{\xi,\zeta}_{\mu\nu\alpha'\beta'}(x,x')$ satisfies
\begin{equation}
\label{tensor_eom}
P_{\xi,\zeta}^{\rho\sigma\mu\nu} G^{\xi,\zeta}_{\mu\nu\alpha'\beta'}(x,x') = g_{\alpha'}^{(\rho} g^{\sigma)}_{\beta'} \delta(x,x') \eqend{.}
\end{equation}
Similar to the vector field, $P_{\xi,\zeta}^{\rho\sigma\mu\nu}$ is not a normally hyperbolic operator for general values of $\xi$ and $\zeta$. However, even in the case $\xi = \zeta = 1$ it is not normally hyperbolic since the coefficient of $\nabla^2$ is not the identity on symmetric rank-2 tensors, $\left( g^{\rho(\mu} g^{\nu)\sigma} - \frac{1}{2} g^{\rho\sigma} g^{\mu\nu} \right) f_{\mu\nu} \neq f^{\rho\sigma}$ (the factor $1/2$ is irrelevant). However, the trace-reversed operator
\begin{equation}\begin{split}
\bar{P}_{1,1}^{\mu\nu\rho\sigma} &\equiv P_{1,1}^{\mu\nu\rho\sigma} - \frac{1}{n-2} g^{\rho\sigma} g_{\alpha\beta} P_{1,1}^{\mu\nu\alpha\beta} \\
&= \frac{1}{2} g^{\rho(\mu} g^{\nu)\sigma} \nabla^2 - R^{\rho(\mu\nu)\sigma}
\end{split}\end{equation}
is normally hyperbolic, and possesses unique retarded and advanced Green's functions $\bar{G}^{1,1}_{\mu\nu\alpha'\beta'}$. From those, one obtains the Green's functions of $P_{1,1}^{\mu\nu\alpha\beta}$ by the same trace reversal, a purely algebraic operation:
\begin{equation}
\label{tensor_green_tracereversed}
G^{1,1}_{\mu\nu\alpha'\beta'} = \bar{G}^{1,1}_{\mu\nu\alpha'\beta'} - \frac{1}{n-2} g_{\mu\nu} g^{\rho\sigma} \bar{G}^{1,1}_{\rho\sigma\alpha'\beta'} \eqend{,}
\end{equation}
and thus their existence (and uniqueness) is also guaranteed. In the literature, it is common to directly use the trace reversed variable $\bar{h}_{\mu\nu} \equiv h_{\mu\nu} - \frac{1}{2} g_{\mu\nu} h$. However, in our case this does not lead to a simplification, and in particular $\bar{P}_{1,1}^{\mu\nu\rho\sigma}$ is \emph{not} the differential operator which one would obtain by replacing $h_{\mu\nu}$ by $\bar{h}_{\mu\nu}$ in the action for $\xi = \zeta = 1$.

\subsubsection*{Trace and divergence identities for \texorpdfstring{$\xi = \zeta = 1$}{xi = zeta = 1}}

To construct the Green's function in the general case, we follow the same strategy as for the vector field. We thus first derive a relation between the divergence of the tensor Green's function and the gradient of vector and scalar ones. Let us introduce a mass parameter
\begin{equation}
\label{mh_def}
\mathfrak{m}^2 \equiv - \frac{4 \Lambda}{n-2} \eqend{.}
\end{equation}
Using that
\begin{equation}
g_{\rho\sigma} P_{1,1}^{\rho\sigma\mu\nu} = - \frac{n-2}{4} g^{\mu\nu} P_{\mathfrak{m}^2} \eqend{,}
\end{equation}
we calculate from Eq.~\eqref{tensor_eom} that
\begin{equation}
P_{\mathfrak{m}^2} \left[ g^{\mu\nu} G^{1,1}_{\mu\nu\alpha'\beta'}(x,x') + \frac{4}{n-2} g_{\alpha'\beta'} G_{\mathfrak{m}^2}(x,x') \right] = 0 \eqend{.}
\end{equation}
Since $P_{\mathfrak{m}^2}$ is a normally hyperbolic operator, the solution of this equation with retarded or advanced boundary conditions is again unique and we infer the trace identity~\cite{lichnerowicz1961}
\begin{equation}
\label{tensor_feyn_trace}
g^{\mu\nu} G^{1,1}_{\mu\nu\alpha'\beta'} = - \frac{4}{n-2} g_{\alpha'\beta'} G_{\mathfrak{m}^2} \eqend{.}
\end{equation}
Using this identity, it follows that
\begin{equation}\begin{split}
&\left( \frac{1}{2} g^{\rho\mu} g^{\nu\sigma} \nabla^2 + R^{\rho\mu\sigma\nu} \right) G^{1,1}_{\mu\nu\alpha'\beta'}(x,x') \\
&\quad= \left( g_{\alpha'}^{(\rho} g^{\sigma)}_{\beta'} - \frac{1}{n-2} g^{\rho\sigma} g_{\alpha'\beta'} \right) \delta(x,x') \eqend{,}
\end{split}\end{equation}
and we calculate
\begin{widetext}
\begin{equation}\begin{split}
&P_{\mathfrak{m}^2/2} \left[ \nabla^\nu G^{1,1}_{\mu\nu\alpha'\beta'} + 2 \nabla_{(\alpha'} G^{\mathfrak{m}^2,1}_{|\mu|\beta')} + \frac{2}{n-2} g_{\alpha'\beta'} \nabla_\mu G_{\mathfrak{m}^2} \right] \\
&\quad= \nabla^\nu \nabla^2 G^{1,1}_{\mu\nu\alpha'\beta'}(x,x') - 2 R^{\rho\nu\sigma}{}_\mu \nabla_\rho G^{1,1}_{\sigma\nu\alpha'\beta'}(x,x') + 2 g_{\mu(\alpha'} \nabla_{\beta')} \delta(x,x') + \frac{2}{n-2} g_{\alpha'\beta'} \nabla_\mu \delta(x,x') \\
&\quad= \nabla_\sigma \left[ 2 g_{\mu(\alpha'} g^\sigma_{\beta')} \delta(x,x') - \frac{2}{n-2} \delta_\mu^\sigma g_{\alpha'\beta'} \delta(x,x') \right] + 2 g_{\mu(\alpha'} \nabla_{\beta')} \delta(x,x') + \frac{2}{n-2} g_{\alpha'\beta'} \nabla_\mu \delta(x,x') = 0 \eqend{.}
\end{split}\end{equation}
\end{widetext}
Since $P_{\mathfrak{m}^2/2}$ is again a normally hyperbolic operator we conclude that~\cite{lichnerowicz1961}
\begin{equation}
\label{tensor_feyn_div}
\nabla^\nu G^{1,1}_{\mu\nu\alpha'\beta'} = - 2 \nabla_{(\alpha'} G^{\mathfrak{m}^2,1}_{|\mu|\beta')} - \frac{2}{n-2} g_{\alpha'\beta'} \nabla_\mu G_{\mathfrak{m}^2} \eqend{.}
\end{equation}
Similar to the vector case, we will also derive this relation as a Ward identity for the free quantum theory in Sec.~\ref{section_ward}. Note that while certain \emph{states} (i.e., Wightman functions or Feynman propagators) for fields of negative mass might be ill-behaved, the retarded/advanced Green's functions are completely well-defined. For example, in de Sitter space where $\Lambda > 0$ the scalar Wightman function is infrared-divergent for the natural Bunch--Davies vacuum state for all $m^2 \leq 0$ (which includes $\mathfrak{m}^2 < 0$), while in the retarded Green's function the problematic infrared divergence cancels out (see, e.g., Ref.~\cite{allen1985}).

\subsubsection*{The graviton Green's function for general \texorpdfstring{$\xi$}{xi} and \texorpdfstring{$\zeta$}{zeta}}

In the general case where either $\xi$ or $\zeta$ (or both) are different from $1$, again inspired from the flat-space Green's function\footnote{\label{footnote_gravprop}See, e.g., Refs.~\cite{capper1980,leonardwoodard2012,froeb2017} and use the formula
\begin{equation*}
\int \mathe^{\mathi p x} (p^2 - \mathi \epsilon)^\alpha \frac{\total^n p}{(2\pi)^n} = \mathi \frac{4^\alpha \Gamma\left( \frac{n}{2}+\alpha \right)}{\pi^\frac{n}{2} \Gamma(-\alpha)} (x^2 + \mathi \epsilon)^{-\alpha-\frac{n}{2}}
\end{equation*}
in $n$ dimensions~\cite{smirnovfeynman}~Eq.~(A.40), \cite{brychkovprudnikov}~Eq.~(8.715) (converted to our conventions).}, we consider the combination
\begin{widetext}
\begin{equation}\begin{split}
\label{graviton_ansatz}
&\tilde{G}^{\xi,\zeta}_{\mu\nu\alpha'\beta'} \equiv G^{\xi,\zeta}_{\mu\nu\alpha'\beta'} + (\xi-1) \left( \nabla_\mu \nabla_{\alpha'} \tilde{G}_{\nu\beta'} + \nabla_\nu \nabla_{\alpha'} \tilde{G}_{\mu\beta'} + \nabla_\mu \nabla_{\beta'} \tilde{G}_{\nu\alpha'} + \nabla_\nu \nabla_{\beta'} \tilde{G}_{\mu\alpha'} \right) \\
&\quad- \frac{4 (1-\zeta)}{(n-2) (1-2\zeta)} \left( g_{\mu\nu} \nabla_{\alpha'} \nabla_{\beta'} + g_{\alpha'\beta'} \nabla_\mu \nabla_\nu \right) \tilde{G}_1 + \frac{4 (1-\zeta)}{(1-2\zeta)^2} \left( (\xi-1) (1-3\zeta) + \frac{n}{n-2} (1-\zeta) \right) \nabla_\mu \nabla_\nu \nabla_{\alpha'} \nabla_{\beta'} \tilde{G}_2
\end{split}\end{equation}
with unknown functions $\tilde{G}_{\nu\beta'}$, $\tilde{G}_1$ and $\tilde{G}_2$. This combination satisfies
\begin{equation}
\label{tensor_xi_p1_tilde_eq1}
P_{1,1}^{\rho\sigma\mu\nu} \tilde{G}^{\xi,\zeta}_{\mu\nu\alpha'\beta'}(x,x') = g_{\alpha'}^{(\rho} g^{\sigma)}_{\beta'} \delta(x,x') \eqend{,}
\end{equation}
if $\tilde{G}_{\nu\beta'}$, $\tilde{G}_1$ and $\tilde{G}_2$ fulfil the conditions
\begin{equation}\begin{split}
\label{tensor_xi_p1_tilde_eqa}
&g_{\alpha'\beta'} \left( g^{\rho\sigma} \nabla^2 - 2 \zeta \nabla^\rho \nabla^\sigma \right) P_{\mathbf{M}^2} \tilde{G}_1 + (n-2) \zeta g^{\rho\sigma} \nabla_{\alpha'} \nabla_{\beta'} P_{\mathbf{M}^2} \tilde{G}_1 - n \frac{1-\zeta}{1-2\zeta} \left( g^{\rho\sigma} \nabla^2 - 2 \zeta \nabla^\rho \nabla^\sigma \right) \nabla_{\alpha'} \nabla_{\beta'} \left( P_{\mathbf{M}^2} \tilde{G}_2 - \tilde{G}_1 \right) \\
&\qquad= \frac{n-2}{2} \zeta \left( g^{\mu\nu} \nabla^\rho \nabla^\sigma + g^{\rho\sigma} \nabla^\mu \nabla^\nu \right) \tilde{G}^{\xi,\zeta}_{\mu\nu\alpha'\beta'} - \frac{n-2}{4} (1+\zeta) g^{\rho\sigma} g^{\mu\nu} \nabla^2 \tilde{G}^{\xi,\zeta}_{\mu\nu\alpha'\beta'}
\end{split}\end{equation}
and
\begin{equation}\begin{split}
\label{tensor_xi_p1_tilde_eqb}
&- 2 \left( \nabla^{(\rho} P_{\mathfrak{m}^2,-\zeta/(1-2\zeta)}^{\sigma)\nu} + \frac{1-2\zeta}{2 \zeta^2} g^{\rho\sigma} P_{\mathbf{M}^2} \nabla^\nu \right) \nabla_{(\alpha'} \tilde{G}_{|\nu|\beta')} - \frac{1-\zeta}{\zeta^2} \left( g^{\rho\sigma} \nabla^2 - 2 \zeta \nabla^\rho \nabla^\sigma \right) \left( \nabla_{\alpha'} \nabla_{\beta'} + \frac{g_{\alpha'\beta'}}{n-2} P_{\mathbf{M}^2} \right) \tilde{G}_1 \\
&\qquad\qquad+ \frac{2}{n-2} \frac{(1-\zeta) [ 1 + (n-3) \zeta ]}{\zeta^2 (1-2\zeta)} \nabla_{\alpha'} \nabla_{\beta'} \left( g^{\rho\sigma} \nabla^2 - 2 \zeta \nabla^\rho \nabla^\sigma \right) \left( P_{\mathbf{M}^2} \tilde{G}_2 - \tilde{G}_1 \right) \\
&\qquad= \nabla^{(\rho} g^{\sigma)(\mu} \nabla^{\nu)} \tilde{G}^{\xi,\zeta}_{\mu\nu\alpha'\beta'} - \frac{1}{2 \zeta} \left( g^{\mu\nu} \nabla^\rho \nabla^\sigma + g^{\rho\sigma} \nabla^\mu \nabla^\nu \right) \tilde{G}^{\xi,\zeta}_{\mu\nu\alpha'\beta'} + \frac{1}{4 \zeta^2} g^{\rho\sigma} g^{\mu\nu} \nabla^2 \tilde{G}^{\xi,\zeta}_{\mu\nu\alpha'\beta'} \eqend{,}
\end{split}\end{equation}
\end{widetext}
where we defined
\begin{equation}
\label{mzeta_def}
\mathbf{M}^2 \equiv - \frac{\zeta}{1-2\zeta} \mathfrak{m}^2 \eqend{.}
\end{equation}
Since $P_{1,1}^{\rho\sigma\mu\nu}$ has unique retarded and advanced Green's functions, we infer that $\tilde{G}^{\xi,\zeta}_{\mu\nu\alpha'\beta'} = G^{1,1}_{\mu\nu\alpha'\beta'}$, and can calculate the right-hand sides of these conditions using the relations~\eqref{vector_xi_div}, \eqref{tensor_feyn_trace} and~\eqref{tensor_feyn_div}. It is then easy to check that Eqns.~\eqref{tensor_xi_p1_tilde_eqa} and~\eqref{tensor_xi_p1_tilde_eqb} are fulfilled if
\begin{subequations}\label{tensor_xi_scalar_eqns}\begin{align}
P_{\mathbf{M}^2} \tilde{G}_1 &= G_{\mathfrak{m}^2} \eqend{,} \\
P_{\mathbf{M}^2} \tilde{G}_2 &= \tilde{G}_1 \eqend{,} \\
g_{\mu\sigma} P_{\mathfrak{m}^2,-\zeta/(1-2\zeta)}^{\sigma\nu} \tilde{G}_{\nu\beta'} &= G^{\mathfrak{m}^2,1}_{\mu\beta'} + \frac{1-\zeta}{\zeta} \nabla_\mu \nabla_{\beta'} \tilde{G}_1 \eqend{.}
\end{align}\end{subequations}
By a straightforward but lengthy calculation, we find a solution of these latter conditions in terms of (mass derivatives of) vector and scalar Green's functions, and replacing those solutions into the ansatz~\eqref{graviton_ansatz}, we find that the retarded or advanced Green's function for the graviton in a general linear covariant gauge takes the form
\begin{widetext}
\begin{equation}\begin{split}
\label{tensor_propagator}
G^{\xi,\zeta}_{\mu\nu\alpha'\beta'} &= G^{1,1}_{\mu\nu\alpha'\beta'} - 2 (\xi-1) \left[ \nabla_{\alpha'} \nabla_{(\mu} \hat{G}^{\mathfrak{m}^2,1}_{\nu)\beta'} + \nabla_{\beta'} \nabla_{(\mu} \hat{G}^{\mathfrak{m}^2,1}_{\nu)\alpha'} \right] - 4 (\xi-1) \frac{1-\zeta}{1-2\zeta} \nabla_\mu \nabla_\nu \nabla_{\alpha'} \nabla_{\beta'} \left( \frac{\hat{G}_{\mathbf{M}^2} - \hat{G}_{\mathfrak{m}^2}}{\mathbf{M}^2 - \mathfrak{m}^2} \right) \\
&\qquad+ \frac{4 (1-\zeta)}{(n-2) (1-2\zeta)} \left( g_{\mu\nu} \nabla_{\alpha'} \nabla_{\beta'} + g_{\alpha'\beta'} \nabla_\mu \nabla_\nu \right) \left( \frac{G_{\mathbf{M}^2} - G_{\mathfrak{m}^2}}{\mathbf{M}^2 - \mathfrak{m}^2} \right) \\
&\qquad+ \frac{4 (1-\zeta)^2}{(1-2\zeta)^2} \left[ \frac{n}{n-2} - (\xi-1) \right] \nabla_\mu \nabla_\nu \nabla_{\alpha'} \nabla_{\beta'} \left[ \frac{G_{\mathbf{M}^2} - G_{\mathfrak{m}^2} - \left( \mathbf{M}^2 - \mathfrak{m}^2 \right) \hat{G}_{\mathbf{M}^2}}{\left( \mathbf{M}^2 - \mathfrak{m}^2 \right)^2} \right] \eqend{.}
\end{split}\end{equation}
\end{widetext}

Lastly, we also want to give expressions for the trace and divergence of the graviton Green's function in the general gauge $\xi, \zeta \neq 1$. Using the relation~\eqref{tensor_feyn_trace} for the trace of the graviton Green's function in the gauge \mbox{$\xi = \zeta = 1$}, the divergence of the vector Green's function~\eqref{vector_feyn_div} and its mass derivative, the equations satisfied by the scalar Green's function~\eqref{scalar_kleingordon} and its mass derivative~\eqref{scalar_kleingordon_massderiv}, we find
\begin{widetext}
\begin{equation}\begin{split}
\label{tensor_general_trace}
g^{\mu\nu} G^{\xi,\zeta}_{\mu\nu\alpha'\beta'} &= \frac{4 \zeta}{(n-2) (1-2\zeta)} \left[ g_{\alpha'\beta'} G_{\mathbf{M}^2} - \frac{n (1-\zeta) - (n-2) (\xi-1) \zeta}{(1-2\zeta)} \nabla_{\alpha'} \nabla_{\beta'} \hat{G}_{\mathbf{M}^2} \right] \eqend{.}
\end{split}\end{equation}
Now using also the relation~\eqref{tensor_feyn_div} for the divergence of the graviton Green's function in the gauge $\xi = \zeta = 1$ and the mass derivative of the vector Eq.~\eqref{vector_eom} we find
\begin{equation}\begin{split}
\label{tensor_general_div}
\nabla^\nu G^{\xi,\zeta}_{\mu\nu\alpha'\beta'} &= - 2 \xi \nabla_{(\alpha'} G^{\mathfrak{m}^2,1}_{|\mu|\beta')} + \frac{2}{(n-2) (1-2\zeta)} g_{\alpha'\beta'} \nabla_\mu G_{\mathbf{M}^2} - \frac{2 \xi (1-\zeta)}{1-2\zeta} \nabla_{\alpha'} \nabla_{\beta'} \nabla_\mu \left( \frac{G_{\mathbf{M}^2} - G_{\mathfrak{m}^2}}{\mathbf{M}^2 - \mathfrak{m}^2} \right) \\
&\quad- \frac{2}{(1-2\zeta)^2} \left[ \frac{n}{n-2} (1-\zeta) - (\xi-1) \zeta \right] \nabla_{\alpha'} \nabla_{\beta'} \nabla_\mu \hat{G}_{\mathbf{M}^2} \eqend{.}
\end{split}\end{equation}
\end{widetext}

\subsubsection*{Special gauges and vanishing cosmological constant}

Let us elaborate here on the form of the graviton Green's function~\eqref{tensor_propagator}, the trace identity~\eqref{tensor_general_trace} and the divergence identity~\eqref{tensor_general_div} for special values of the gauge parameters and for the case with a vanishing cosmological constant.

\ $\xi \to 1$:\quad This limit is clearly seen to be regular for all three identities and will make some of the terms in those expressions vanish.

\ $\zeta \to 1$:\quad While the trace identity \eqref{tensor_general_trace} is clearly regular in this limit, the regularity of the other two identities is not apparent, since from the definition~\eqref{mzeta_def} we have $\mathbf{M}^2 \to \mathfrak{m}^2$ for $\zeta \to 1$. However, the mass terms always appear in the combination
\begin{equation}
\frac{1-\zeta}{\mathbf{M}^2 - \mathfrak{m}^2} = - \frac{1-2\zeta}{\mathfrak{m}^2} \to \frac{1}{\mathfrak{m}^2} \eqend{,}
\end{equation}
which remains regular as $\zeta \to 1$, and it follows that the Green's function is given by
\begin{widetext}
\begin{equation}
\label{tensor_propagator_zeta1}
G^{\xi,1}_{\mu\nu\alpha'\beta'} = G^{1,1}_{\mu\nu\alpha'\beta'} - 2 (\xi-1) \left( \nabla_{\alpha'} \nabla_{(\mu} \hat{G}^{\mathfrak{m}^2,1}_{\nu)\beta'} + \nabla_{\beta'} \nabla_{(\mu} \hat{G}^{\mathfrak{m}^2,1}_{\nu)\alpha'} \right) \eqend{,}
\end{equation}
while trace and divergence are
\begin{subequations}\begin{align}
g^{\mu\nu} G^{\xi,1}_{\mu\nu\alpha'\beta'} &= - \frac{4}{n-2} g_{\alpha'\beta'} G_{\mathfrak{m}^2} + 4 (\xi-1) \nabla_{\alpha'} \nabla_{\beta'} \hat{G}_{\mathfrak{m}^2} \eqend{,} \\
\nabla^\nu G^{\xi,1}_{\mu\nu\alpha'\beta'} &= - 2 \xi \nabla_{(\alpha'} G^{\mathfrak{m}^2,1}_{|\mu|\beta')} - \frac{2}{n-2} g_{\alpha'\beta'} \nabla_\mu G_{\mathfrak{m}^2} + 2 (\xi-1) \nabla_{\alpha'} \nabla_{\beta'} \nabla_\mu \hat{G}_{\mathfrak{m}^2} \eqend{.}
\end{align}\end{subequations}
\end{widetext}

\ $\xi \to 1$, $\zeta \to 1$:\quad From the above expressions, $G^{1,1}_{\mu\nu\alpha'\beta'}$ and the identities~\eqref{tensor_feyn_trace} and~\eqref{tensor_feyn_div} are clearly recovered in this limit.

\ $\zeta \to 0$:\quad In this limit, $\mathbf{M} \to 0$ and the trace of the Green's function vanishes
\begin{equation}
g^{\mu\nu} G^{\xi,0}_{\mu\nu\alpha'\beta'} = 0 \eqend{.}
\end{equation}

\ $\xi \to 0$, $\zeta \to \infty$:\quad In this limit, $\mathbf{M} \to \mathfrak{m}^2/2$ and the trace term in the gauge-fixing action~\eqref{gauge-fixing-action} disappears. We have
\begin{subequations}\begin{align}
g^{\mu\nu} G^{0,\infty}_{\mu\nu\alpha'\beta'} &= - \frac{2}{n-2} \left( g_{\alpha'\beta'} G_{\mathfrak{m}^2/2} - \nabla_{\alpha'} \nabla_{\beta'} \hat{G}_{\mathfrak{m}^2/2} \right) \eqend{,} \\
\nabla^\nu G^{0,\infty}_{\mu\nu\alpha'\beta'} & = 0 \eqend{.}
\end{align}\end{subequations}
In view of vanishing of the divergence of Green's function, we can call this gauge generalized Landau gauge.

\ $\xi \to 0$, $\zeta = n/2$:\quad In this gauge, the Green's function decomposes into two parts
\begin{equation}\begin{split}
G^{0,n/2}_{\mu\nu\alpha'\beta'} = G^\text{TT}_{\mu\nu\alpha'\beta'} - \frac{2}{(n-2) (n-1)} g_{\mu\nu} g_{\alpha'\beta'} G_{\mathbf{M}^2} \eqend{,}
\end{split}\end{equation}
where the ``spin-2'' part
\begin{equation}\begin{split}
&G^\text{TT}_{\mu\nu\alpha'\beta'} = G^{1,1}_{\mu\nu\alpha'\beta'} + 2 \left[ \nabla_{\alpha'} \nabla_{(\mu} \hat{G}^{\mathfrak{m}^2,1}_{\nu)\beta'} + \nabla_{\beta'} \nabla_{(\mu} \hat{G}^{\mathfrak{m}^2,1}_{\nu)\alpha'} \right] \\
&\quad+ \frac{2}{n-1} \left( g_{\mu\nu} \nabla_{\alpha'} \nabla_{\beta'} + g_{\alpha'\beta'} \nabla_\mu \nabla_\nu \right) \left( \frac{G_{\mathbf{M}^2} - G_{\mathfrak{m}^2}}{\mathbf{M}^2 - \mathfrak{m}^2} \right) \\
&\quad+ \frac{2 (n-2)}{n-1} \nabla_\mu \nabla_\nu \nabla_{\alpha'} \nabla_{\beta'} \left[ \frac{G_{\mathbf{M}^2} - G_{\mathfrak{m}^2}}{\left( \mathbf{M}^2 - \mathfrak{m}^2 \right)^2} - \frac{\hat{G}_{\mathfrak{m}^2}}{\mathbf{M}^2 - \mathfrak{m}^2} \right] \\
&\quad+ \frac{2}{(n-2) (n-1)} g_{\alpha'\beta'} g_{\mu\nu} G_{\mathbf{M}^2}
\end{split}\end{equation}
is transverse and traceless,
\begin{equation}
\nabla^\mu G^\text{TT}_{\mu\nu\alpha'\beta'} = 0 = g^{\mu\nu} G^\text{TT}_{\mu\nu\alpha'\beta'} \eqend{,}
\end{equation}
as can be checked using the trace~\eqref{tensor_general_trace} and divergence identities~\eqref{tensor_general_div}. This gauge is used in AdS/CFT calculations~\cite{dhokeretal1999}.

\ $\Lambda \to 0$:\quad At first sight, this limit might seem divergent since both $\mathbf{M}^2$ and $\mathfrak{m}^2$ vanish in this limit. To show that the limit actually exists, we expand around zero mass
\begin{subequations}\begin{align}
G_{\mathfrak{m}^2} &= G_0 + \mathfrak{m}^2 \hat{G}_0 + \frac{1}{2} \mathfrak{m}^4 \dhat{G}_0 + \bigo{\mathfrak{m}^6} \eqend{,} \\
\hat{G}_{\mathfrak{m}^2} &= \hat{G}_0 + \mathfrak{m}^2 \dhat{G}_0 + \bigo{\mathfrak{m}^4} \eqend{,}
\end{align}\end{subequations}
and the analogous equations for $\mathbf{M}^2$, and obtain [using also the definitions of $\mathfrak{m}^2$~\eqref{mh_def} and $\mathbf{M}^2$~\eqref{mzeta_def}]
\begin{subequations}\begin{align}
\frac{G_{\mathbf{M}^2} - G_{\mathfrak{m}^2}}{\mathbf{M}^2 - \mathfrak{m}^2} &\to \hat{G}_0 \eqend{,} \\
\frac{\hat{G}_{\mathfrak{m}^2}}{\mathbf{M}^2 - \mathfrak{m}^2} &\to \frac{\hat{G}_0}{\mathbf{M}^2 - \mathfrak{m}^2} - \frac{1-2\zeta}{1-\zeta} \dhat{G}_0 \eqend{,} \\
\frac{\hat{G}_{\mathbf{M}^2}}{\mathbf{M}^2 - \mathfrak{m}^2} &\to \frac{\hat{G}_0}{\mathbf{M}^2 - \mathfrak{m}^2} + \frac{\zeta}{1-\zeta} \dhat{G}_0 \eqend{,} \\
\frac{G_{\mathbf{M}^2} - G_{\mathfrak{m}^2}}{\left( \mathbf{M}^2 - \mathfrak{m}^2 \right)^2} &\to \frac{\hat{G}_0}{\mathbf{M}^2 - \mathfrak{m}^2} - \frac{1-3\zeta}{2 (1-\zeta)} \dhat{G}_0 \eqend{.}
\end{align}\end{subequations}
The potentially divergent terms cancel out in the full Green's function, and it follows that
\begin{widetext}
\begin{equation}\begin{split}
\label{tensor_propagator_zeromass}
G^{\xi,\zeta}_{\mu\nu\alpha'\beta'} \Big\rvert_{\Lambda = 0} &= G^{1,1}_{\mu\nu\alpha'\beta'} \Big\rvert_{\Lambda = 0} - 2 (\xi-1) \left( \nabla_{\alpha'} \nabla_{(\mu} \hat{G}^{0,1}_{\nu)\beta'} + \nabla_{\beta'} \nabla_{(\mu} \hat{G}^{0,1}_{\nu)\alpha'} \right) + \frac{4 (1-\zeta)}{(n-2) (1-2\zeta)} \left( g_{\mu\nu} \nabla_{\alpha'} \nabla_{\beta'} + g_{\alpha'\beta'} \nabla_\mu \nabla_\nu \right) \hat{G}_0 \\
&\quad- \frac{2 (1-\zeta)}{(1-2\zeta)^2} \Big[ \frac{n}{n-2} (1-\zeta) + (\xi-1) (1-3\zeta) \Big] \nabla_\mu \nabla_\nu \nabla_{\alpha'} \nabla_{\beta'} \dhat{G}_0
\end{split}\end{equation}
and
\begin{equation}\begin{split}
\nabla^\nu G^{\xi,\zeta}_{\mu\nu\alpha'\beta'} \Big\rvert_{\Lambda = 0} &= - 2 \xi \nabla_{(\alpha'} G^{0,1}_{|\mu|\beta')} + \frac{2}{(n-2) (1-2\zeta)} g_{\alpha'\beta'} \nabla_\mu G_0 \\
&\quad- \frac{2}{(n-2) (1-2\zeta)^2} \left[ 2 [ (n-1) - (n-2) \zeta ] (1-\zeta) + (n-2) (\xi-1) (1-4\zeta+2\zeta^2) \right] \nabla_{\alpha'} \nabla_{\beta'} \nabla_\mu \hat{G}_0 \eqend{.}
\end{split}\end{equation}
\end{widetext}
In flat space where in addition $R_{\mu\nu\rho\sigma} = 0$, this coincides with well-known results (see footnote~\ref{footnote_gravprop} on page~\pageref{footnote_gravprop}).

\section{Ward identities}
\label{section_ward}

As mentioned in the previous sections, the relations~\eqref{vector_xi_div}, \eqref{tensor_general_trace} and~\eqref{tensor_general_div} can be derived as Ward identities of the free quantum theory. For the divergence relations~\eqref{vector_xi_div} and~\eqref{tensor_general_div}, this has already been noted previously~\cite{allenottewill1992,hollands2008,belokognefolacci2016}, and in this section we show how to derive all three identities in the framework of BRST quantization of gauge theories. A mathematically rigorous formulation of BRST quantization in the algebraic approach to quantum field theory on curved spacetimes has been given in Refs.~\cite{hollands2008,fredenhagenrejzner2013,brunettifredenhagenreijzner2016,taslimitehrani2017}, but we only need some basic facts which we state in the following.

\subsection{Massive vector boson}

In the BRST formalism, one introduces in addition to the vector field $A_\mu$ fermionic ghost and antighost fields $c$ and $\bar{c}$, and an auxiliary bosonic (Nakanishi-Lautrup) field $B$. For a massive (Proca) vector field, the theory is not a gauge theory, and one must add an additional auxiliary scalar field $\phi$, known as Stueckelberg field~\cite{stueckelberg1938a,stueckelberg1938b,belokognefolacci2016}. The full action then reads
\begin{equation}
S = \frac{1}{2} \int \bigg[ A_\mu P_{m^2,\xi}^{\mu\nu} A_\nu + \phi P_{\xi m^2} \phi - 2 \bar{c} P_{\xi m^2} c + \xi \tilde{B}^2 \bigg] \total_g x
\end{equation}
with the differential operators $P_{m^2,\xi}^{\mu\nu}$ and $P_{\xi m^2}$ defined in Eqns.~\eqref{vector_p_def} and~\eqref{scalar_p_def}, and where we defined
\begin{equation}
\tilde{B} \equiv B + \xi^{-1} \nabla_\mu A^\mu + m \phi \eqend{,}
\end{equation}
which completely decouples from the rest of the theory. In the massless limit, we recover the gauge-fixed action of a free gauge vector in the BRST formulation, and we can also obtain the original massive theory in the gauge $\phi = 0$. Furthermore, in the limit $\xi \to \infty$ in which the differential operator $P_{m^2,\xi}^{\mu\nu}$ gives the Proca operator~\eqref{vector_proca}, both the scalar and the ghosts become infinitely massive and drop out of the physical spectrum, such that the Proca theory is recovered.

It is now straightforward to check that $S$ is invariant under the action of the free (linearized) BRST differential $\brst_0$, which is a nilpotent ($\brst_0^2 = 0$) fermionic differential acting according to
\begin{subequations}\label{brst0_action_ym}\begin{align}
\brst_0 A_\mu &= \nabla_\mu c \eqend{,} \\
\brst_0 c &= 0 \eqend{,} \\
\brst_0 \bar{c} &= B \eqend{,} \\
\brst_0 B &= 0 \eqend{,} \\
\brst_0 \phi &= - m c \eqend{.}
\end{align}\end{subequations}
We see that $\brst_0$ generalizes the gauge symmetry to include the additional fields, replacing the gauge parameter with the ghost field. In the interacting classical theory, i.e., the full non-Abelian Yang-Mills(-Higgs) theory, the full nonlinear BRST differential $\brst$ extends this action while remaining nilpotent, and in the quantum theory, one needs to extend it further to a quantum BRST differential $\mathsf{q} = \brst + \bigo{\hbar}$~\cite{froebhollandhollands2016,taslimitehrani2017}. The observables in the quantum theory are then the interacting fields and composite operators invariant under the action of $\mathsf{q}$, but here we restrict to the free quantum theory and only need $\brst_0$. A general Ward identity then reads
\begin{equation}\begin{split}
\label{ward_identity}
0 &= \bra{\psi} \brst_0 \mathcal{T} \mathcal{O}_1 \cdots \mathcal{O}_n \ket{\psi} \\
&= \sum_{i=1}^n (-1)^{\sum_{j=1}^{i-1} \epsilon_j} \bra{\psi} \mathcal{T} \mathcal{O}_1 \cdots \left( \brst_0 \mathcal{O}_i \right) \cdots \mathcal{O}_n \ket{\psi} \eqend{,}
\end{split}\end{equation}
where $\ket{\psi}$ is a Hadamard state, $\mathcal{T}$ is time-ordering, $\mathcal{O}_i$ are field operators (including composite operators, which in the free theory are local and covariant Wick powers defined with respect to a Hadamard parametrix), and $\epsilon_i$ is the Grassmann parity of $\mathcal{O}_i$.

The free quantum fields have the following Feynman propagators in a Hadamard state $\ket{\psi}$:
\begin{subequations}\label{ward_feynman}\begin{align}
\bra{\psi} \mathcal{T} A_\mu(x) A_{\rho'}(x') \ket{\psi} &= \mathi G^{\text{F},m^2,\xi}_{\mu\rho'}(x,x') \eqend{,} \\
\bra{\psi} \mathcal{T} \phi(x) \phi(x') \ket{\psi} &= \mathi G^\text{F}_{\xi m^2}(x,x') \eqend{,} \\
\bra{\psi} \mathcal{T} c(x) \bar{c}(x') \ket{\psi} &= - \mathi G^\text{F}_{\xi m^2}(x,x') \label{ward_feynman_ghost} \eqend{,} \\
\bra{\psi} \mathcal{T} \tilde{B}(x) \tilde{B}(x') \ket{\psi} &= \frac{\mathi}{\xi} \delta(x,x') \eqend{,}
\end{align}\end{subequations}
and analogously the Wightman functions $G^+$ are given by the same formulas without the time-ordering $\mathcal{T}$. The retarded Green's functions can then be recovered from these according to the relation~\eqref{scalar_retarded}: $G^\text{ret}(x,x') = G^\text{F}(x,x') - G^+(x',x)$. Alternatively, one can consider the purely algebraic covariant (anti-)commutators
\begin{subequations}\label{ward_commutator}\begin{align}
[ A_\mu(x), A_{\rho'}(x') ] &= \mathi \Delta^{m^2,\xi}_{\mu\rho'}(x,x') \eqend{,} \\
[ \phi(x), \phi(x') ] &= \mathi \Delta_{\xi m^2}(x,x') \eqend{,} \\
\{ c(x), \bar{c}(x') \} &= - \mathi \Delta_{\xi m^2}(x,x') \eqend{,} \\
[ \tilde{B}(x), \tilde{B}(x') ] &= 0 \eqend{,}
\end{align}\end{subequations}
where $\Delta(x,x') \equiv G^\text{adv}(x,x') - G^\text{ret}(x,x')$ is called the causal propagator or Pauli--Jordan function. This would be more in line with the spirit of the algebraic approach to quantum field theory, where one constructs the algebra of free or interacting quantum fields first (including renormalization), and worries about states and expectation values afterwards. Let us again denote the right-hand sides by an unspecified ``$G$''.

We then use the identity
\begin{equation}
\bra{\psi} \brst_0 \mathcal{T} A_\mu(x) \bar{c}(x') \ket{\psi} = 0
\end{equation}
(or the commutator of those fields), which gives
\begin{equation}\begin{split}
0 &= \bra{\psi} \mathcal{T} \brst_0 A_\mu(x) \bar{c}(x') \ket{\psi} + \bra{\psi} \mathcal{T} A_\mu(x) \brst_0 \bar{c}(x') \ket{\psi} \\
&= \nabla_\mu \bra{\psi} \mathcal{T} c(x) \bar{c}(x') \ket{\psi} \\
&\qquad+ \bra{\psi} \mathcal{T} A_\mu(x) \left( \tilde{B} - \xi^{-1} \nabla_{\rho'} A^{\rho'} - m \phi \right)(x') \ket{\psi} \\
&= - \mathi \xi^{-1} \left[ \xi \nabla_\mu G_{\xi m^2}(x,x') + \nabla^{\rho'} G^{m^2,\xi}_{\mu\rho'}(x,x') \right] \eqend{,}
\end{split}\end{equation}
which is relation~\eqref{vector_xi_div}. Note that derivatives are taken outside the time-ordered product, which is in accordance with the algebraic approach~\cite{hollandswald2001,hollandswald2002}, and with path integral calculations (where it is sometimes called $\mathcal{T}^*$ product). One might also wonder if it is necessary for the scalar $\phi$ and the ghosts to have the same propagator~\eqref{ward_feynman}; this follows by expanding the Ward identity $0 = \bra{\psi} \brst_0 \mathcal{T} \phi(x) \bar{c}(x') \ket{\psi}$.

\subsection{Graviton}

For the graviton, both ghost and antighost as well as the auxiliary field obtain a Lorentz index, and the free BRST differential acts as
\begin{subequations}\label{brst0_action_grav}\begin{align}
\brst_0 h_{\mu\nu} &= \nabla_\mu c_\nu + \nabla_\nu c_\mu \eqend{,} \\
\brst_0 c_\mu &= 0 \eqend{,} \\
\brst_0 \bar{c}_\mu &= B_\mu \eqend{,} \\
\brst_0 B_\mu &= 0 \eqend{.}
\end{align}\end{subequations}
The BRST-extended action reads
\begin{equation}\begin{split}
\label{graviton_action_brst}
S &= \frac{1}{2} \int h_{\rho\sigma} P_{\xi,\zeta}^{\rho\sigma\mu\nu} h_{\mu\nu} \total_g x + \frac{\xi}{2} \int \tilde{B}^\mu \tilde{B}_\mu \total_g x \\
&\quad- \int \bar{c}_\mu P_{\mathfrak{m}^2,-\zeta/(1-2\zeta)}^{\mu\nu} c_\nu \total_g x
\end{split}\end{equation}
with
\begin{equation}
\tilde{B}_\mu \equiv B_\mu + \frac{1}{\xi} \left( \nabla^\nu h_{\mu\nu} - \frac{1}{2 \zeta} \nabla_\mu h \right) \eqend{,}
\end{equation}
the mass parameter $\mathfrak{m}^2$ from Eq.~\eqref{mh_def} and the differential operator $P_{\xi,\zeta}^{\rho\sigma\mu\nu}$ of Eq.~\eqref{tensor_p_def}. Again it is straightforward to check that $\brst_0 S = 0$. The free quantum fields have the following Feynman propagators in a Hadamard state $\ket{\psi}$:
\begin{subequations}\label{ward_feynman_grav}\begin{align}
\bra{\psi} \mathcal{T} h_{\mu\nu}(x) h_{\rho'\sigma'}(x') \ket{\psi} &= \mathi G^{\text{F},\xi,\zeta}_{\mu\nu\rho'\sigma'}(x,x') \eqend{,} \\
\bra{\psi} \mathcal{T} c_\mu(x) \bar{c}_{\rho'}(x') \ket{\psi} &= - \mathi G^{\text{F},\mathfrak{m}^2,-\zeta/(1-2\zeta)}_{\mu\rho'}(x,x') \eqend{,} \label{ward_feynman_grav_ghost} \\
\bra{\psi} \mathcal{T} \tilde{B}_\mu(x) \tilde{B}_{\rho'}(x') \ket{\psi} &= \frac{\mathi}{\xi} g_{\mu\rho'} \delta(x,x') \eqend{,}
\end{align}\end{subequations}
where the propagator~\eqref{ward_feynman_grav_ghost} is the vector one with the specified mass and gauge parameter, and the analogous expressions for the commutators. To derive the divergence relation~\eqref{tensor_general_div}, we expand the Ward identity
\begin{equation}
\bra{\psi} \brst_0 \mathcal{T} h_{\mu\nu}(x) \bar{c}_{\rho'}(x') \ket{\psi} = 0
\end{equation}
and obtain
\begin{widetext}
\begin{equation}\begin{split}
\label{tensor_general_divward}
0 &= \bra{\psi} \mathcal{T} \left( \nabla_\mu c_\nu + \nabla_\nu c_\mu \right)(x) \bar{c}_{\rho'}(x') \ket{\psi} + \bra{\psi} \mathcal{T} h_{\mu\nu}(x) \left[ \tilde{B}_{\rho'} - \frac{1}{\xi} \left( \nabla^{\sigma'} h_{\rho'\sigma'} - \frac{1}{2 \zeta} \nabla_{\rho'} h \right) \right](x') \ket{\psi} \\
&= - \mathi \xi^{-1} \left[ \nabla^{\sigma'} G^{\xi,\zeta}_{\mu\nu\rho'\sigma'}(x,x') - \frac{1}{2 \zeta} g^{\alpha'\beta'} \nabla_{\rho'} G^{\xi,\zeta}_{\mu\nu\alpha'\beta'}(x,x') + 2 \xi \nabla_{(\mu} G^{\mathfrak{m}^2,-\zeta/(1-2\zeta)}_{\nu)\rho'}(x,x') \right] \eqend{.}
\end{split}\end{equation}
\end{widetext}
Using the trace identity~\eqref{tensor_general_trace} to replace the second term and the relation~\eqref{vector_propagator} between the vector Green's function in different gauges, the divergence identity~\eqref{tensor_general_div} follows.

Since the gauge symmetry of the free classical theory $h_{\mu\nu} \to h_{\mu\nu} + 2 \nabla_{(\mu} v_{\nu)}$ arises from linearized diffeomorphisms, the divergence identity in the form~\eqref{tensor_general_divward} is the Ward identity associated to diffeomorphisms (and the $c_\mu$ are the diffeomorphism ghosts). Similarly, the trace identity~\eqref{tensor_general_trace} would be the Ward identity associated to linearized Weyl transformations $h_{\mu\nu} \to h_{\mu\nu} + g_{\mu\nu} w$, but this is not a (gauge) symmetry of the original action $S^{(0)}$~\eqref{graviton_action}. To derive it, we thus again need to consider an extended theory with an additional compensating scalar field $\phi$, which transforms under linearized Weyl transformations as $\phi \to \phi + (n-2) w$, and whose action is obtained from the original action $S^{(0)}$~\eqref{graviton_action} by replacing $h_{\mu\nu} \to h_{\mu\nu} - 1/(n-2) g_{\mu\nu} \phi$. This theory is now invariant under linearized Weyl transformations, and to gauge fix this new symmetry we add a (Weyl) ghost $d$, antighost $\bar{d}$, and auxiliary field $F$. The gauge-fixed extended action $S'$ then reads
\begin{equation}\begin{split}
S' &= \frac{1}{2} \int h_{\rho\sigma} P_{\xi,\zeta}^{\rho\sigma\mu\nu} h_{\mu\nu} \total_g x + \frac{\xi}{2} \int \tilde{B}_\mu \tilde{B}^\mu \total_g x \\
&\quad+ \int \tilde{F} P_{\mathbf{M}^2} \phi \total_g x + \frac{1}{2} \mathcal{M}^2 \int \phi^2 \total_g x \\
&\quad- \int \bar{c}_\mu P_{\mathfrak{m}^2,-\zeta/(1-2\zeta)}^{\mu\nu} c_\nu \total_g x - \int \bar{d} P_{\mathbf{M}^2} d \total_g x \\
&\quad+ \frac{n - 2 \zeta - (n-2) \zeta \xi}{2\zeta} \int \bar{c}^\mu \nabla_\mu d \total_g x \eqend{,}
\end{split}\end{equation}
where now
\begin{subequations}\begin{align}
\tilde{B}_\mu &\equiv B_\mu + \frac{1}{\xi} \left( \nabla^\nu h_{\mu\nu} - \frac{1}{2\zeta} \nabla_\mu h \right) + \frac{1}{2} \nabla_\mu \phi \eqend{,} \\
\tilde{F} &\equiv \frac{F}{n-2} - \frac{1-2\zeta}{4 \zeta} h - \frac{2 (n-1) - (n-2) \xi}{8 (n-2)} \phi \eqend{,} \\
\mathcal{M}^2 &\equiv \frac{n - 2 \zeta - (n-2) \xi \zeta}{4 (n-2) (1-2\zeta)} \mathfrak{m}^2 \eqend{.}
\end{align}\end{subequations}
The free BRST differential acts according to
\begin{subequations}\label{brst0_action_grav2}\begin{align}
\brst_0 h_{\mu\nu} &= \nabla_\mu c_\nu + \nabla_\nu c_\mu + g_{\mu\nu} d \eqend{,} \\
\brst_0 \phi &= (n-2) d \eqend{,} \\
\brst_0 d &= 0 \eqend{,} \\
\brst_0 \bar{d} &= F \eqend{,} \\
\brst_0 F &= 0 \eqend{,}
\end{align}\end{subequations}
with the action on $\bar{c}_\mu$, $c_\mu$ and $B_\mu$ unchanged from~\eqref{brst0_action_grav}, and again $\brst_0 S' = 0$. The Feynman propagators for $h_{\mu\nu}$ and $\tilde{B}$ are unchanged, while the others now read
\begin{subequations}\begin{align}
\bra{\psi} \mathcal{T} \phi(x) \phi(x') \ket{\psi} &= 0 \eqend{,} \\
\bra{\psi} \mathcal{T} \tilde{F}(x) \phi(x') \ket{\psi} &= \mathi G^\text{F}_{\mathbf{M}^2}(x,x') \eqend{,} \\
\bra{\psi} \mathcal{T} \tilde{F}(x) \tilde{F}(x') \ket{\psi} &= - \mathi \mathcal{M}^2 \hat{G}^\text{F}_{\mathbf{M}^2}(x,x') \eqend{,} \\
\bra{\psi} \mathcal{T} c_\mu(x) \bar{c}_{\rho'}(x') \ket{\psi} &= - \mathi G^{\text{F},\mathfrak{m}^2,-\zeta/(1-2\zeta)}_{\mu\rho'}(x,x') \eqend{,} \\
\begin{split}
\bra{\psi} \mathcal{T} c_\mu(x) \bar{d}(x') \ket{\psi} &= \mathi \frac{n - 2 \zeta - (n-2) \zeta \xi}{2 (1-2\zeta)} \\
&\qquad\times \nabla^\mu \hat{G}^\text{F}_{\mathbf{M}^2}(x,x') \eqend{,}
\end{split} \\
\bra{\psi} \mathcal{T} d(x) \bar{d}(x') \ket{\psi} &= - \mathi G^\text{F}_{\mathbf{M}^2}(x,x') \eqend{,} \\
\bra{\psi} \mathcal{T} d(x) \bar{c}_{\rho'}(x') \ket{\psi} &= 0 \eqend{.}
\end{align}\end{subequations}
To derive the trace relation~\eqref{tensor_general_trace}, we expand the identity
\begin{equation}
\bra{\psi} \brst_0 \mathcal{T} h_{\mu\nu}(x) \bar{d}(x') \ket{\psi} = 0
\end{equation}
and obtain
\begin{widetext}
\begin{equation}\begin{split}
0 &= \bra{\psi} \mathcal{T} \left( 2 \nabla_{(\mu} c_{\nu)} + g_{\mu\nu} d \right)(x) \bar{d}(x') \ket{\psi} + (n-2) \bra{\psi} \mathcal{T} h_{\mu\nu}(x) \left[ \tilde{F} + \frac{1-2\zeta}{4 \zeta} h + \frac{2 (n-1) - (n-2) \xi}{8 (n-2)} \phi \right](x') \ket{\psi} \\
&= \mathi \left[ \frac{(n-2) (1-2\zeta)}{4 \zeta} g^{\rho'\sigma'} G^{\xi,\zeta}_{\mu\nu\rho'\sigma'}(x,x') + \frac{n - 2 \zeta - (n-2) \zeta \xi}{1-2\zeta} \nabla_\mu \nabla_\nu \hat{G}_{\mathbf{M}^2}(x,x') - g_{\mu\nu} G_{\mathbf{M}^2}(x,x') \right] \eqend{,}
\end{split}\end{equation}
\end{widetext}
which coincides with Eq.~\eqref{tensor_general_trace}.

While we had previously derived the divergence and trace identities only for the (retarded or advanced) Green's functions, their derivation as Ward identities means that they also must hold for Feynman propagators, and in general for the state-dependent correlation functions if the theory is to be consistent. This is obviously a much stronger requirement, and further complicates the already intricate issue of the construction of Hadamard states in general curved spacetimes. However, once states have been found in one particular gauge of the family of linear and covariant gauges that we study in this paper, one can obtain states that fulfil the divergence and trace identities for the whole family by relating the Wightman functions or Feynman propagators in the same way as the Green's functions~\eqref{vector_propagator}, \eqref{vector_propagator_massless} and~\eqref{tensor_propagator}.

\section{Hadamard expansion}
\label{section_hadamard}

As explained in the introduction, the Wightman function for a scalar field in any Hadamard state in four dimensions has the form~\cite{hadamard1932,baerginouxpfaeffle2007}
\begin{equation}
\label{wightman_4d_scalar2}
G^+_{m^2}(x,x') = H^+_{m^2}(x,x') - \frac{\mathi}{8 \pi^2} W_{m^2}(x,x')
\end{equation}
for $x'$ in a normal geodesic neighborhood of $x$, where $H^+$ is the Hadamard parametrix~\eqref{hadamard_4d_scalar} with the Wightman prescription\footnote{The prefactor is again a matter of convention.} $\sigma_\epsilon = \sigma + \mathi \epsilon (t-t')$, and where $U_{m^2}$, $V_{m^2}$ and $W_{m^2}$ are smooth biscalars. Since $G^+_{m^2}$ is a bisolution of the Klein--Gordon equation, it follows that $H^+_{m^2}$ is a bisolution up to a smooth remainder:
\begin{equation}\begin{split}
P_{m^2}(x) H^+_{m^2}(x,x') &= f(x,x') \eqend{,} \\
P_{m^2}(x') H^+_{m^2}(x,x') &= f'(x,x') \eqend{,}
\end{split}\end{equation}
with $f$ and $f'$ being (unspecified) smooth functions. In fact, this is the definition of a parametrix for a general differential operator~\cite{baerginouxpfaeffle2007,baerginoux2012}. The Hadamard parametrix $H^\text{F}_{m^2}$ for the Feynman propagator is obtained from Eq.~\eqref{hadamard_4d_scalar} using the Feynman prescription $\sigma_\epsilon = \sigma + \mathi \epsilon$. This parametrix is a bisolution modulo a smooth remainder to the inhomogeneous equation:
\begin{equation}\begin{split}
P_{m^2}(x) H^\text{F}_{m^2}(x,x') &= \delta(x,x') + f(x,x') \eqend{,} \\
P_{m^2}(x') H^\text{F}_{m^2}(x,x') &= \delta(x,x') + f'(x,x') \eqend{,}
\end{split}\end{equation}
where again $f$ and $f'$ are smooth functions. Similarly, one can define advanced, retarded, Dyson (anti-Feynman) Hadamard parametrices, which coincide with the respective propagators/Green's functions up to a smooth remainder in any normal geodesic neighborhood. Those involve the same biscalars $U_{m^2}$ and $V_{m^2}$, but differ in the type of $\mathi \epsilon$ prescription needed near $\sigma = 0$ to properly define them as distributions, and one can use the relations~\eqref{sokhotski_plemelj} to relate them. In the following, we will work with the general form~\eqref{hadamard_4d_scalar}, and the analogue for vector and tensor fields, and for notational convenience drop the subscript $\epsilon$ on $\sigma$. Furthermore, we will denote a generic Green's function or two-point function by $G$.

\subsection{Scalar field}

It is well known that the biscalars $U_{m^2}$, $V_{m^2}$ and $W_{m^2}$ possess an asymptotic expansion as $x \to x'$, of the form
\begin{subequations}\label{hadamard_4d_scalarexpansion}\begin{align}
U_{m^2} &= U_{m^2}^{(0)} \eqend{,} \\
\{ V/W \}_{m^2} &= \sum_{k=0}^\infty \{ V/W \}_{m^2}^{(k)} \sigma^k
\end{align}\end{subequations}
with smooth biscalars $U_{m^2}^{(0)}$, $V_{m^2}^{(k)}$ and $W_{m^2}^{(k)}$, which for analytic spacetimes is even convergent (see, e.g., Refs.~\cite{moretti2000,hollandswald2005,decaninifolacci2008,baerginouxpfaeffle2007} and references therein). By requiring $G^+_{m^2}$ to solve the Klein--Gordon equation outside of coincidence and comparing manifest powers of $\sigma$, one obtains
\begin{equation}
\label{hadamard_4d_u0}
U_{m^2}^{(0)} = \sqrt{\Delta}
\end{equation}
with the van~Vleck--Morette determinant~\cite{dewittbrehme1960,synge1960}
\begin{equation}
\Delta(x,x') = - \left[ g(x) g(x') \right]^{-\frac{1}{2}} \det\left[ \nabla_\alpha \nabla_{\beta'} \sigma(x,x') \right] \eqend{,}
\end{equation}
and the recursion relations
\begin{subequations}\label{hadamard_4d_scalar_recursion}\begin{align}
Q_{2k+4} V_{m^2}^{(k+1)} &= - \frac{1}{k+1} P_{m^2} V_{m^2}^{(k)} \eqend{,} \\
Q_{2k+4} W_{m^2}^{(k+1)} &= - \frac{1}{k+1} \left( P_{m^2} W_{m^2}^{(k)} + Q_{4k+6} V_{m^2}^{(k+1)} \right)
\end{align}\end{subequations}
with the differential operator
\begin{equation}
\label{qk_def}
Q_k \equiv 2 \nabla^\mu \sigma \nabla_\mu - \nabla^\mu \sigma \nabla_\mu \ln \Delta + k \eqend{,}
\end{equation}
subject to the boundary condition
\begin{equation}
\label{hadamard_4d_v0_bdy}
Q_2 V_{m^2}^{(0)} = - P_{m^2} \sqrt{\Delta} \eqend{.}
\end{equation}
Again it is seen that $U_{m^2}$ and $V_{m^2}$ are completely determined geometrically, while for $W_{m^2}$ the first coefficient is an arbitrary solution of the free Klein--Gordon equation
\begin{equation}
P_{m^2} W_{m^2}^{(0)} = 0 \eqend{,}
\end{equation}
which encodes the state-dependence of the two-point function. Imposing smoothness, there is a unique solution to the recursion relations~\eqref{hadamard_4d_scalar_recursion} for which the coefficients are symmetric~\cite{moretti2000}. This solution can be given explicitly as an integral in Riemannian normal coordinates, but in the following we only need that the unique smooth solution to $Q_k f = 0$ is $f = 0$, from which it follows in particular that the $V_{m^2}^{(k)}$ are polynomials of order $k+1$ in $m^2$~\cite{moretti2000,decaninifolacci2006}. For completeness, we give an explicit solution in Appendix~\ref{appendix_riemann}.

\subsubsection*{Mass derivatives}

Since the Green's functions for vector and tensor fields in a general gauge~\eqref{vector_propagator}, \eqref{vector_propagator_massless} and~\eqref{tensor_propagator} also involve mass derivatives, we need to calculate the corresponding coefficients of the Hadamard expansion. As explained previously, certain states (i.e., Wightman functions or Feynman propagators) might be ill-behaved for certain ranges of the mass parameter, and the same applies to their mass derivatives. However, this problem does not arise for the retarded or advanced propagators (which are state-independent), and is thus confined to the $W$ biscalar. In the following, we will also present formulas for the coefficients $W_{m^2}^{(k)}$, with the understanding that we only consider such states for which the resulting expressions are well-defined. In contrast, the corresponding formulas for $U_{m^2}^{(0)}$ and the $V_{m^2}^{(k)}$ are always well-defined; in fact, $U_{m^2}^{(0)}$~\eqref{hadamard_4d_u0} is mass-independent, and as stated before $V_{m^2}^{(k)}$ is a polynomial in $m^2$ of order $k+1$. The Hadamard expansion of the mass derivative~\eqref{scalar_massderivative} is then simply obtained by taking a mass derivative of Eqns.~\eqref{wightman_scalar}, \eqref{wightman_4d_scalar2} or~\eqref{hadamard_4d_scalar} and the recursion relations~\eqref{hadamard_4d_scalar_recursion} for the coefficients. Since $U_{m^2}^{(0)}$ is mass-independent, it follows that
\begin{equation}
\label{hadamard_4d_scalar_massderiv}
\hat{G}_{m^2} = - \frac{\mathi}{8 \pi^2} \left[ \hat{V}_{m^2} \ln\left( \mu^2 \sigma \right) + \hat{W}_{m^2} \right] \eqend{,}
\end{equation}
where $\hat{V}$ and $\hat{W}$ have the asymptotic expansions
\begin{equation}
\hat{V}_{m^2} = \sum_{k=0}^\infty \hat{V}_{m^2}^{(k)}\sigma^k \eqend{,} \qquad \hat{W}_{m^2} = \sum_{k=0}^\infty \hat{W}_{m^2}^{(k)} \sigma^k \eqend{.}
\end{equation}
They fulfil the recursion relations
\begin{subequations}\label{hadamard_4d_scalar_massder_recursion}\begin{align}
Q_{2k+4} \hat{V}_{m^2}^{(k+1)} &= - \frac{1}{k+1} \left( P_{m^2} \hat{V}_{m^2}^{(k)} - V_{m^2}^{(k)} \right) \eqend{,} \\
\begin{split}
Q_{2k+4} \hat{W}_{m^2}^{(k+1)} &= - \frac{1}{k+1} \left( P_{m^2} \hat{W}_{m^2}^{(k)} - W_{m^2}^{(k)} \right) \\
&\quad- \frac{1}{k+1} Q_{4k+6} \hat{V}_{m^2}^{(k+1)}
\end{split}
\end{align}\end{subequations}
with the boundary condition
\begin{equation}
\label{hadamard_4d_hatv_bdy}
\hat{V}_{m^2}^{(0)} = \frac{1}{2} \sqrt{\Delta}
\end{equation}
for $\hat{V}_{m^2}^{(0)}$, and $\hat{W}_{m^2}^{(0)}$ fulfilling $P_{m^2} \hat{W}_{m^2}^{(0)} = W_{m^2}^{(0)}$.

For later use, we now show by induction that for all $k \geq 0$
\begin{equation}
\label{hadamard_4d_hatv_in_v}
\hat{V}_{m^2}^{(k+1)} = \frac{1}{2(k+1)} V_{m^2}^{(k)} \eqend{.}
\end{equation}
Take first $k = 0$, which by the recursion relation~\eqref{hadamard_4d_scalar_massder_recursion} and the boundary conditions~\eqref{hadamard_4d_v0_bdy} and~\eqref{hadamard_4d_hatv_bdy} fulfils
\begin{equation}\begin{split}
Q_4 \hat{V}_{m^2}^{(1)} &= - P_{m^2} \hat{V}_{m^2}^{(0)} + V_{m^2}^{(0)} \\
&= \frac{1}{2} Q_2 V_{m^2}^{(0)} + V_{m^2}^{(0)} = \frac{1}{2} Q_4 V_{m^2}^{(0)} \eqend{.}
\end{split}\end{equation}
Therefore,
\begin{equation}
Q_4 \left( \hat{V}_{m^2}^{(1)} - \frac{1}{2} V_{m^2}^{(0)} \right) = 0 \eqend{,}
\end{equation}
and the unique solution of this first-order differential equation which is smooth vanishes~\cite{moretti2000}. Assume now that $k \geq 1$, and that the relation~\eqref{hadamard_4d_hatv_in_v} has been shown up to order $k-1$. Applying the differential operator $Q_{2k+4}$ on Eq.~\eqref{hadamard_4d_hatv_in_v} and using Eq.~\eqref{hadamard_4d_scalar_massder_recursion}, we obtain
\begin{equation}\begin{split}
&Q_{2k+4} \left[ \hat{V}_{m^2}^{(k+1)} - \frac{1}{2(k+1)} V_{m^2}^{(k)} \right] \\
&\quad= - \frac{1}{2(k+1)} \left( Q_{2k+4} V_{m^2}^{(k)} + 2 P_{m^2} \hat{V}_{m^2}^{(k)} - 2 V_{m^2}^{(k)} \right) \eqend{.}
\end{split}\end{equation}
By the induction hypothesis we have $\hat{V}_{m^2}^{(k)} = 1/(2k) V_{m^2}^{(k-1)}$, which using Eq.~\eqref{hadamard_4d_scalar_recursion} leads to
\begin{equation}\begin{split}
&Q_{2k+4} \left[ \hat{V}_{m^2}^{(k+1)} - \frac{1}{2(k+1)} V_{m^2}^{(k)} \right] \\
&\quad= - \frac{1}{2(k+1)} \left( Q_{2k+2} V_{m^2}^{(k)} + \frac{1}{k} P_{m^2} V_{m^2}^{(k-1)} \right) = 0 \eqend{.}
\end{split}\end{equation}
Again, the unique smooth solution of this first-order differential equation vanishes, and we obtain Eq.~\eqref{hadamard_4d_hatv_in_v}. However, no similar relation exists for the $\hat{W}_{m^2}^{(k)}$.

Similarly, for the second mass derivatives we have
\begin{equation}
\dhat{G}_{m^2} = - \frac{\mathi}{8 \pi^2} \left[ \dhat{V}_{m^2} \ln \left( \mu^2 \sigma \right) + \dhat{W}_{m^2} \right] \eqend{,}
\end{equation}
with (since $\hat{V}_{m^2}^{(0)}$ is mass-independent, the first coefficient $\dhat{V}_{m^2}^{(0)}$ vanishes)
\begin{equation}
\dhat{V}_{m^2} = \sum_{k=1}^\infty \dhat{V}_{m^2}^{(k)} \sigma^k \eqend{,} \qquad \dhat{W}_{m^2} = \sum_{k=0}^\infty \dhat{W}_{m^2}^{(k)} \sigma^k \eqend{,}
\end{equation}
the recursion relations
\begin{subequations}\label{hadamard_4d_scalar_massder2_recursion}\begin{align}
Q_{2k+4} \dhat{V}_{m^2}^{(k+1)} &= - \frac{1}{k+1} \left( P_{m^2} \dhat{V}_{m^2}^{(k)} - 2 \hat{V}_{m^2}^{(k)} \right) \eqend{,} \\
\begin{split}
Q_{2k+4} \dhat{W}_{m^2}^{(k+1)} &= - \frac{1}{k+1} \left( P_{m^2} \dhat{W}_{m^2}^{(k)} - 2 \hat{W}_{m^2}^{(k)} \right) \\
&\quad- \frac{1}{k+1} Q_{4k+6} \dhat{V}_{m^2}^{(k+1)}
\end{split}
\end{align}\end{subequations}
with the boundary condition
\begin{equation}
\label{hadamard_4d_dhatv_bdy}
\dhat{V}_{m^2}^{(1)} = \frac{1}{4} \sqrt{\Delta} \eqend{,}
\end{equation}
and $\dhat{W}_{m^2}^{(0)}$ fulfilling
\begin{equation}
P_{m^2} \dhat{W}_{m^2}^{(0)} = 2 \hat{W}_{m^2}^{(0)} \eqend{.}
\end{equation}
Moreover, taking a mass derivative of Eq.~\eqref{hadamard_4d_hatv_in_v} we obtain for all $k \geq 1$
\begin{equation}
\label{hadamard_4d_dhatv_in_v}
\dhat{V}_{m^2}^{(k+1)} = \frac{1}{2(k+1)} \hat{V}_{m^2}^{(k)} = \frac{1}{4 k (k+1)} V_{m^2}^{(k-1)} \eqend{.}
\end{equation}

Since the coefficients $V_{m^2}^{(k)}$ are polynomials in $m^2$, one can also derive formulas which relate coefficients for different masses. The easiest way to obtain those is to expand around zero mass
\begin{equation}
\label{hadamard_4d_v_massexpansion_general}
V_{m^2}^{(k)} = \sum_{\ell=0}^{k+1} \frac{1}{\ell!} (m^2)^\ell \left[ \frac{\partial^\ell}{\partial (m^2)^\ell} V_{m^2}^{(k)} \right]_{m^2 = 0} \eqend{,}
\end{equation}
and then use the relations~\eqref{hadamard_4d_hatv_in_v} and~\eqref{hadamard_4d_dhatv_in_v} and their generalizations to higher mass derivatives, together with the boundary conditions~\eqref{hadamard_4d_v0_bdy}, \eqref{hadamard_4d_hatv_bdy} and~\eqref{hadamard_4d_dhatv_bdy} and their generalizations to higher mass derivatives. Later on we will need these expressions for $k=0,1,2$, where we obtain
\begin{subequations}\label{hadamard_4d_v_massexpansion}\begin{align}
V_{m^2}^{(0)} &= V_0^{(0)} + m^2 \hat{V}_0^{(0)} = V_0^{(0)} + \frac{1}{2} m^2 \sqrt{\Delta} \eqend{,} \\
\begin{split}
V_{m^2}^{(1)} &= V_0^{(1)} + \frac{1}{2} m^2 V_0^{(0)} + \frac{1}{8} m^4 \sqrt{\Delta} \\
&= V_0^{(1)} + \frac{1}{2} m^2 V_{m^2}^{(0)} - \frac{1}{8} m^4 \sqrt{\Delta} \eqend{,}
\end{split} \\
V_{m^2}^{(2)} &= V_0^{(2)} + \frac{1}{4} m^2 V_0^{(1)} + \frac{1}{16} m^4 V_0^{(0)} + \frac{1}{96} m^6 \sqrt{\Delta} \eqend{.}
\end{align}\end{subequations}

\subsection{Vector field}

The vector Green's function and the local Hadamard parametrix in Feynman gauge $\xi = 1$ have been studied quite extensively in the literature, see e.g. Refs.~\cite{dewittbrehme1960, sahlmannverch2001,fewsterpfenning2003,dappiaggisiemssen2013,finsterstrohmaier2015,gerardwrochna2015}. In this gauge, the Hadamard expansion takes the form~\cite{dewittbrehme1960}
\begin{equation}
\label{hadamard_4d_vector}
G^{m^2,1}_{\nu\beta'} = - \frac{\mathi}{8 \pi^2} \left[ \frac{U^{m^2,1}_{\nu\beta'}}{\sigma} + V^{m^2,1}_{\nu\beta'} \ln\left( \mu^2 \sigma \right) + W^{m^2,1}_{\nu\beta'} \right] \eqend{,}
\end{equation}
where the same assertions as in the scalar case apply. In particular, the functions $U$, $V$ and $W$ are smooth symmetric bitensors possessing an asymptotic expansion of the form
\begin{subequations}\begin{align}
U^{m^2,1}_{\nu\beta'} &= U^{m^2,1(0)}_{\nu\beta'} \eqend{,} \\
V^{m^2,1}_{\nu\beta'} &= \sum_{k=0}^\infty V^{m^2,1(k)}_{\nu\beta'} \sigma^k \eqend{,} \\
W^{m^2,1}_{\nu\beta'} &= \sum_{k=0}^\infty W^{m^2,1(k)}_{\nu\beta'} \sigma^k \eqend{,}
\end{align}\end{subequations}
and imposing the equation of motion~\eqref{vector_eom} outside of coincidence and comparing manifest powers of $\sigma$, they fulfil the recursion relations
\begin{subequations}\label{hadamard_4d_vector_vw_recursion}\begin{align}
Q_{2k+4} V^{m^2,1(k+1)}_{\nu\beta'} &= - \frac{1}{k+1} g_{\nu\mu} P_{m^2,1}^{\mu\rho} V^{m^2,1(k)}_{\rho\beta'} \eqend{,} \\
\begin{split}
Q_{2k+4} W^{m^2,1(k+1)}_{\nu\beta'} &= - \frac{1}{k+1} g_{\nu\mu} P_{m^2,1}^{\mu\rho} W^{m^2,1(k)}_{\rho\beta'} \\
&\quad- \frac{1}{k+1} Q_{4k+6} V^{m^2,1(k+1)}_{\nu\beta'}
\end{split}
\end{align}\end{subequations}
with the boundary conditions
\begin{subequations}\label{hadamard_4d_vector_uv_bdy}\begin{align}
U^{m^2,1(0)}_{\nu\beta'} &= \sqrt{\Delta} \, g_{\nu\beta'} \eqend{,} \\
Q_2 V^{m^2,1(0)}_{\nu\beta'} &= - g_{\nu\mu} P_{m^2,1}^{\mu\rho} \left( \sqrt{\Delta} \, g_{\rho\beta'} \right) \eqend{,}
\end{align}\end{subequations}
and $W^{m^2,1(0)}_{\nu\beta'}$ being an arbitrary smooth solution of the equation of motion $P_{m^2,1}^{\mu\rho} W^{m^2,1(0)}_{\rho\beta'} = 0$, encoding the state dependence. An explicit solution in Riemannian normal coordinates can be given in the same way as for the scalar field. Here again appears the parallel propagator $g_{\mu\beta'}$, which is defined as the unique solution to~\cite{synge1960}
\begin{equation}
\label{bitensor_parallel_props}
\nabla^\rho \sigma \nabla_\rho g_{\mu\beta'} = 0 \eqend{,} \qquad \lim_{x' \to x} g_{\mu\beta'} = g_{\mu\beta} \eqend{.}
\end{equation}

For a general gauge, the retarded or advanced Green's functions are given by Eq.~\eqref{vector_propagator} in the massive case and Eq.~\eqref{vector_propagator_massless} in the massless case, which completely determines the $U$ and $V$ coefficients and thus the Hadamard parametrix; we assume that the relation between the Wightman function or Feynman propagator in different gauges is also given by Eqns.~\eqref{vector_propagator} and~\eqref{vector_propagator_massless}, which then also determines the $W$ coefficients. Using the Hadamard expansion of the scalar propagator~\eqref{hadamard_4d_scalar} and taking into account that the first coefficient $U_{m^2}^{(0)}$ is mass-independent, it then follows that their Hadamard expansion is given by
\begin{equation}\begin{split}
\label{hadamard_4d_vector_xi}
G^{m^2,\xi}_{\nu\beta'}(x,x') &= - \frac{\mathi}{8 \pi^2} \Bigg[ \frac{U^{m^2,\xi(-1)}_{\nu\beta'}}{\sigma^2} + \frac{U^{m^2,\xi(0)}_{\nu\beta'}}{\sigma} \\
&\qquad\qquad+ V^{m^2,\xi}_{\nu\beta'} \ln \left( \mu^2 \sigma \right) + W^{m^2,\xi}_{\nu\beta'} \Bigg]
\end{split}\end{equation}
with the asymptotic expansions
\begin{equation}
V^{m^2,\xi}_{\nu\beta'} = \sum_{k=0}^\infty V^{m^2,\xi(k)}_{\nu\beta'} \sigma^k \eqend{,} \quad W^{m^2,\xi}_{\nu\beta'} = \sum_{k=0}^\infty W^{m^2,\xi(k)}_{\nu\beta'} \sigma^k \eqend{.}
\end{equation}
The coefficients $\{U/V/W\}^{m^2,\xi(k)}_{\nu\beta'}$ are obtained by inserting the expansions~\eqref{hadamard_4d_vector} and~\eqref{wightman_scalar} into Eq.~\eqref{vector_propagator}, performing the derivatives and comparing manifest powers of $\sigma$. Using also the mass expansion~\eqref{hadamard_4d_v_massexpansion}, this gives
\begin{subequations}\label{hadamard_4d_vector_uvw}\begin{align}
U^{m^2,\xi(-1)}_{\nu\beta'} &= \frac{\xi-1}{2} \sqrt{\Delta} \, \sigma_\nu \sigma_{\beta'} \eqend{,} \\
\begin{split}
U^{m^2,\xi(0)}_{\nu\beta'} &= U^{m^2,1}_{\nu\beta'} - \frac{\xi-1}{2} \bigg[ \left( 2 \sigma_{(\nu} \nabla_{\beta')} + \sigma_{\nu\beta'} \right) \sqrt{\Delta} \\
&\hspace{8em}+ \frac{1}{2} \left( V_{m^2}^{(0)} + V_{\xi m^2}^{(0)} \right) \sigma_\nu \sigma_{\beta'} \bigg] \eqend{,}
\end{split} \\
\begin{split}
V^{m^2,\xi(k)}_{\nu\beta'} &= V^{m^2,1(k)}_{\nu\beta'} - \nabla_\nu \nabla_{\beta'} \frac{V_{\xi m^2}^{(k)} - V_{m^2}^{(k)}}{m^2} \\
&\quad- \frac{k+1}{m^2} \bigg[ (k+2) \sigma_\nu \sigma_{\beta'} \left( V_{\xi m^2}^{(k+2)} - V_{m^2}^{(k+2)} \right) \\
&\qquad+ \left( 2 \sigma_{(\nu} \nabla_{\beta')} + \sigma_{\nu\beta'} \right) \left( V_{\xi m^2}^{(k+1)} - V_{m^2}^{(k+1)} \right) \bigg] \eqend{,}
\end{split}
\end{align}\end{subequations}
where to shorten the resulting expressions we have defined
\begin{equation}
\sigma_{\mu \cdots \nu} \equiv \nabla_\nu \cdots \nabla_\mu \sigma \eqend{,}
\end{equation}
and the lengthy expressions for the $W$ coefficients are given in Appendix~\ref{appendix_w}.

We note the appearance of a term proportional to $\sigma^{-2}$ in the Hadamard expansion~\eqref{hadamard_4d_vector_xi}, which seems more singular than the $\sigma^{-1}$ term. However, the presence of $\sigma_\nu \sigma_{\beta'}$ in the numerator of this term reduces the strength of the singularity, which in a mathematically precise way is captured by the scaling degree or degree of divergence (scaling degree minus spacetime dimension)~\cite{steinmann1971,weinstein1978}. Near coincidence, we have $\sigma(x,x') \approx (x-x')^2/2$ with the Minkowski squared distance $(x-x')^2$, which upon a rescaling $\{x/x'\} \to \lambda \{x/x'\}$ is rescaled by a factor $\lambda^2$, and $\sigma^{-1}$ has thus scaling degree $2$ and degree of divergence $2 - 4 = -2$. Since the degree of divergence is negative, $\sigma^{-1}$ (with any $\epsilon$ prescription) is already well defined also at coincidence, i.e., the limit $\epsilon \to 0$ exists after smearing with arbitrary test functions whose support contains $x = x'$ (see, e.g., Refs.~\cite{brunettifredenhagen2000,hollandswald2002,hollands2008}). Similarly, $\sigma_\nu(x,x') \approx (x-x')_\nu$ near coincidence which scales with a factor $\lambda$. Therefore, the degree of divergence of $\sigma_\nu \sigma_{\beta'}/\sigma^2$ is also $-2 < 0$, and this term is again already well defined at coincidence for any $\epsilon$ prescription.

One might still wonder why this case is different from the Hadamard expansion of the product $[ G_{m^2}(x,x') ]^2$, where also terms proportional to $\sigma^{-2}$ appear. There, such terms must in general be renormalized depending on the concrete type of Green's function/Wightman function/propagator. For example, for the Feynman prescription $\sigma + \mathi \epsilon$ we have $( \sigma + \mathi \epsilon )^{-2}$, which is \emph{not} a well-defined distribution in four dimensions. However, in our case this term comes from taking a derivative of $\sigma^{-1}$, which is well-defined for any prescription as a distributional derivative. For example, for the Feynman prescription $\sigma + \mathi \epsilon$ we take a derivative of the relation~\eqref{sokhotski_plemelj} and obtain
\begin{equation}
- \frac{\total}{\total \sigma} \frac{1}{\sigma + \mathi \epsilon} = \mathcal{P}\!f \frac{1}{\sigma^2} + \mathi \pi \delta'(\sigma) \eqend{,}
\end{equation}
where the right-hand side is a well-defined distribution in four dimensions, and the left-hand side is how $\sigma^{-2}$ in the Hadamard expansion~\eqref{hadamard_4d_vector_xi} (with Feynman prescription) should be understood. The higher negative powers of $\sigma$ which appear for the graviton are defined analogously.

The massless limit can be taken easily using the expansion~\eqref{hadamard_4d_v_massexpansion_general} (and its analogue for the $W$ coefficients, which we assume to exist as explained above). Using also the relations~\eqref{hadamard_4d_hatv_bdy} and~\eqref{hadamard_4d_hatv_in_v} we obtain
\begin{subequations}\begin{align}
U^{0,\xi(-1)}_{\nu\beta'} &= \frac{\xi-1}{2} \sqrt{\Delta} \, \sigma_\nu \sigma_{\beta'} \eqend{,} \\
\begin{split}
U^{0,\xi(0)}_{\nu\beta'} &= U^{0,1}_{\nu\beta'} - \frac{\xi-1}{2} \bigg[ \left( 2 \sigma_{(\nu} \nabla_{\beta')} + \sigma_{\nu\beta'} \right) \sqrt{\Delta} \\
&\hspace{8em}+ V_0^{(0)} \sigma_\nu \sigma_{\beta'} \bigg] \eqend{,}
\end{split} \\
\begin{split}
V^{0,\xi(k)}_{\nu\beta'} &= V^{0,1(k)}_{\nu\beta'} - \frac{\xi-1}{2} \bigg[ \nabla_\nu \nabla_{\beta'} \hat{V}_0^{(k)} \\
&\qquad+ (k+1) \sigma_\nu \sigma_{\beta'} V_0^{(k+1)} \\
&\qquad+ \left( 2 \sigma_{(\nu} \nabla_{\beta')} + \sigma_{\nu\beta'} \right) V_0^{(k)} \bigg] \eqend{,}
\end{split}
\end{align}\end{subequations}
and the expressions for the $W$ coefficients are again given in Appendix~\ref{appendix_w}. Of course, these expressions are identical to the ones that would be obtained by inserting the Hadamard expansion of $\hat{G}_0$~\eqref{hadamard_4d_scalar_massderiv} into the massless vector propagator~\eqref{vector_propagator_massless}.

\subsubsection*{Mass derivatives}

For use in the graviton case, we also need the mass derivative of the vector coefficients. In Feynman gauge $\xi = 1$, we take a mass derivative of the recursion relations~\eqref{hadamard_4d_vector_vw_recursion} and obtain
\begin{subequations}\label{hadamard_4d_vector_vw_massder_recursion}\begin{align}
\begin{split}
&Q_{2k+4} \hat{V}^{m^2,1(k+1)}_{\nu\beta'} \\
&\qquad= - \frac{1}{k+1} \left( g_{\nu\mu} P_{m^2,1}^{\mu\rho} \hat{V}^{m^2,1(k)}_{\rho\beta'} - V^{m^2,1(k)}_{\nu\beta'} \right) \eqend{,}
\end{split} \\
\begin{split}
&Q_{2k+4} \hat{W}^{m^2,1(k+1)}_{\nu\beta'} = - \frac{1}{k+1} \bigg( g_{\nu\mu} P_{m^2,1}^{\mu\rho} \hat{W}^{m^2,1(k)}_{\rho\beta'} \\
&\hspace{6em}- W^{m^2,1(k)}_{\nu\beta'} + Q_{4k+6} \hat{V}^{m^2,1(k+1)}_{\nu\beta'} \bigg) \eqend{,}
\end{split}
\end{align}\end{subequations}
and taking a mass derivative of the boundary conditions~\eqref{hadamard_4d_vector_uv_bdy} we have
\begin{equation}
\label{hadamard_4d_vector_uv_massder_bdy}
\hat{U}^{m^2,1(0)}_{\nu\beta'} = 0 \eqend{,} \qquad Q_2 \hat{V}^{m^2,1(0)}_{\nu\beta'} = \sqrt{\Delta} \, g_{\nu\beta'} \eqend{.}
\end{equation}
The last equation again admits a unique smooth solution [which can be checked using the properties of the bitensor of parallel transport~\eqref{bitensor_parallel_props}], given by
\begin{equation}
\label{hadamard_4d_vector_v_massder_bdy}
\hat{V}^{m^2,1(0)}_{\nu\beta'} = \frac{1}{2} \sqrt{\Delta} \, g_{\nu\beta'} \eqend{,}
\end{equation}
and similar to the case of the scalar field the boundary condition for $\hat{W}^{m^2,1(0)}_{\nu\beta'}$ is
\begin{equation}
g_{\nu\mu} P_{m^2,1}^{\mu\rho} \hat{W}^{m^2,1(0)}_{\rho\beta'} = W^{m^2,1(0)}_{\nu\beta'} \eqend{.}
\end{equation}
In complete analogy to the scalar case, we show by induction that for all $k \geq 0$
\begin{equation}
\label{hadamard_4d_vector_hatv_in_v}
\hat{V}^{m^2,1(k+1)}_{\nu\beta'} = \frac{1}{2(k+1)} V^{m^2,1(k)}_{\nu\beta'} \eqend{.}
\end{equation}
Take first $k = 0$, which by the recursion relation~\eqref{hadamard_4d_vector_vw_massder_recursion} and the boundary conditions~\eqref{hadamard_4d_vector_uv_bdy} and~\eqref{hadamard_4d_vector_v_massder_bdy} fulfils
\begin{equation}\begin{split}
Q_4 \hat{V}^{m^2,1(1)}_{\nu\beta'} &= - g_{\nu\mu} P_{m^2,1}^{\mu\rho} \hat{V}^{m^2,1(0)}_{\rho\beta'} + V^{m^2,1(0)}_{\nu\beta'} \\
&= \frac{1}{2} Q_2 V^{m^2,1(0)}_{\nu\beta'} + V^{m^2,1(0)}_{\nu\beta'} = \frac{1}{2} Q_4 V^{m^2,1(0)}_{\nu\beta'} \eqend{.}
\end{split}\end{equation}
The unique smooth solution to this first-order differential equation is given by~\eqref{hadamard_4d_vector_hatv_in_v}. Assume now that $k \geq 1$, and that the relation~\eqref{hadamard_4d_vector_hatv_in_v} has been shown up to order $k-1$. Applying the differential operator $Q_{2k+4}$ and using equation~\eqref{hadamard_4d_vector_vw_massder_recursion}, we obtain
\begin{equation}\begin{split}
&Q_{2k+4} \left[ 2 (k+1) \hat{V}^{m^2,1(k+1)}_{\nu\beta'} - V^{m^2,1(k)}_{\nu\beta'} \right] \\
&\quad= - Q_{2k+4} V^{m^2,1(k)}_{\nu\beta'} - 2 g_{\nu\mu} P_{m^2,1}^{\mu\rho} \hat{V}^{m^2,1(k)}_{\rho\beta'} + 2 V^{m^2,1(k)}_{\nu\beta'} \eqend{.}
\end{split}\end{equation}
Since by induction we may assume that $\hat{V}^{m^2,1(k)}_{\rho\beta'} = 1/(2k) V^{m^2,1(k-1)}_{\rho\beta'}$, it follows that
\begin{equation}\begin{split}
&Q_{2k+4} \left[ 2 (k+1) \hat{V}^{m^2,1(k+1)}_{\nu\beta'} - V^{m^2,1(k)}_{\nu\beta'} \right] \\
&\quad= - Q_{2k+2} V^{m^2,1(k)}_{\nu\beta'} - \frac{1}{k} g_{\nu\mu} P_{m^2,1}^{\mu\rho} V^{m^2,1(k-1)}_{\rho\beta'} = 0
\end{split}\end{equation}
using the recursion relation~\eqref{hadamard_4d_vector_vw_recursion}, and the unique smooth solution is again given by~\eqref{hadamard_4d_vector_hatv_in_v}. Again, no similar relation exists for the $\hat{W}^{m^2,1(k)}_{\rho\beta'}$.

In a general gauge, the mass derivative of the vector coefficients is easily computed by taking a mass derivative of the general Hadamard expansion~\eqref{hadamard_4d_vector_xi}. Since many terms are mass-independent, their derivative vanishes and we obtain
\begin{equation}
\label{hadamard_4d_vector_xi_massderiv}
\hat{G}^{m^2,\xi}_{\nu\beta'} = - \frac{\mathi}{8 \pi^2} \left[ \frac{\hat{U}^{m^2,\xi(0)}_{\nu\beta'}}{\sigma} + \hat{V}^{m^2,\xi}_{\nu\beta'} \ln \left( \mu^2 \sigma \right) + \hat{W}^{m^2,\xi}_{\nu\beta'} \right]
\end{equation}
with the asymptotic expansions
\begin{equation}
\hat{V}^{m^2,\xi}_{\nu\beta'} = \sum_{k=0}^\infty \hat{V}^{m^2,\xi(k)}_{\nu\beta'} \sigma^k \eqend{,} \qquad \hat{W}^{m^2,\xi}_{\nu\beta'} = \sum_{k=0}^\infty \hat{W}^{m^2,\xi(k)}_{\nu\beta'} \sigma^k \eqend{.}
\end{equation}
Using the relations~\eqref{hadamard_4d_vector_v_massder_bdy} and~\eqref{hadamard_4d_vector_hatv_in_v} for the mass derivative of the Feynman gauge vector coefficients, \eqref{hadamard_4d_hatv_bdy} and~\eqref{hadamard_4d_hatv_in_v} for the mass derivative of scalar coefficients, and~\eqref{hadamard_4d_v_massexpansion} for different masses, one can compute these coefficients, and we delegate the lengthy expressions to Appendix~\ref{appendix_w}.

Let us here make a remark on the singular nature of the Hadamard expansion of the vector propagator. Naively, one might expect that the Hadamard expansion~\eqref{hadamard_4d_vector_xi} of the vector Green's function in the general gauge $\xi \neq 1$ contains a term proportional to $\sigma^{-3}$, arising from two derivatives acting on the term proportional to $\sigma^{-1}$ in the Hadamard expansion~\eqref{hadamard_4d_scalar} of the scalar Green's function. However, since the general gauge vector Green's function~\eqref{vector_propagator} involves the difference between two scalar Green's functions with different masses, and the coefficient $U^{(0)}_{m^2}$ of this term is independent of the mass~\eqref{hadamard_4d_u0}, this term actually vanishes, and the most singular term in the Hadamard expansion~\eqref{hadamard_4d_vector_xi} is proportional to $\sigma^{-2}$. Its coefficient $U^{m^2,\xi(-1)}$ is again mass-independent~\eqref{hadamard_4d_vector_uvw}, and thus the Hadamard expansion of the mass derivative of the vector Green's function~\eqref{hadamard_4d_vector_xi_massderiv} has only $\sigma^{-1}$ as its most singular term.

\subsection{Tensor field}
In the analogue of Feynman gauge $\xi = \zeta = 1$, we have the Hadamard expansion~\cite{allenfolacciottewill1988,minosasakitanaka1997}
\begin{equation}
\label{hadamard_4d_tensor}
G^{1,1}_{\mu\nu\alpha'\beta'} = - \frac{\mathi}{8 \pi^2} \left[ \frac{U^{1,1}_{\mu\nu\alpha'\beta'}}{\sigma} + V^{1,1}_{\mu\nu\alpha'\beta'} \ln\left( \mu^2 \sigma \right) + W^{1,1}_{\mu\nu\alpha'\beta'} \right] \eqend{,}
\end{equation}
where the same assertions as in the scalar and vector case apply. In particular, the functions $U$, $V$ and $W$ are smooth symmetric bitensors possessing an asymptotic expansion of the form
\begin{subequations}\begin{align}
U^{1,1}_{\mu\nu\alpha'\beta'} &= U^{1,1(0)}_{\mu\nu\alpha'\beta'} \eqend{,} \\
V^{1,1}_{\mu\nu\alpha'\beta'} &= \sum_{k=0}^\infty V^{1,1(k)}_{\mu\nu\alpha'\beta'} \sigma^k \eqend{,} \\
W^{1,1}_{\mu\nu\alpha'\beta'} &= \sum_{k=0}^\infty W^{1,1(k)}_{\mu\nu\alpha'\beta'} \sigma^k \eqend{.}
\end{align}\end{subequations}
Requiring $G^{1,1}_{\mu\nu\alpha'\beta'}$ to be a solution of the equation of motion $P_{1,1}^{\mu\nu\rho\sigma} G^{1,1}_{\rho\sigma\alpha'\beta'} = 0$ outside of coincidence, they fulfil the recursion relations
\begin{subequations}\label{hadamard_4d_tensor_vw_recursion}\begin{align}
Q_{2k+4} V^{1,1(k+1)}_{\mu\nu\alpha'\beta'} &= - \frac{1}{k+1} g_{\mu\rho} g_{\nu\sigma} P_{1,1}^{\rho\sigma\kappa\lambda} V^{1,1(k)}_{\kappa\lambda\alpha'\beta'} \eqend{,} \\
\begin{split}
Q_{2k+4} W^{1,1(k+1)}_{\mu\nu\alpha'\beta'} &= - \frac{1}{k+1} g_{\mu\rho} g_{\nu\sigma} P_{1,1}^{\rho\sigma\kappa\lambda} W^{1,1(k)}_{\kappa\lambda\alpha'\beta'} \\
&\quad- \frac{1}{k+1} Q_{4k+6} V^{1,1(k+1)}_{\mu\nu\alpha'\beta'}
\end{split}
\end{align}\end{subequations}
with the boundary conditions
\begin{subequations}\label{hadamard_4d_tensor_uv_bdy}\begin{align}
U^{1,1(0)}_{\mu\nu\alpha'\beta'} &= \sqrt{\Delta} \, \left( g_{\alpha'(\mu} g_{\nu)\beta'} - \frac{1}{2} g_{\mu\nu} g_{\alpha'\beta'} \right) \eqend{,} \\
Q_2 V^{1,1(0)}_{\mu\nu\alpha'\beta'} &= - g_{\mu\rho} g_{\nu\sigma} P_{1,1}^{\rho\sigma\kappa\lambda} U^{1,1(0)}_{\kappa\lambda\alpha'\beta'} \eqend{,}
\end{align}\end{subequations}
and $W^{1,1(0)}_{\mu\nu\alpha'\beta'}$ is an arbitrary smooth solution of the equation of motion $P_{1,1}^{\mu\nu\rho\sigma} W^{1,1(0)}_{\rho\sigma\alpha'\beta'} = 0$.

To obtain the Hadamard expansion in the general gauge $\xi, \zeta \neq 1$ we have to insert the expansions~\eqref{wightman_scalar}, \eqref{hadamard_4d_scalar_massderiv} and~\eqref{hadamard_4d_vector_xi_massderiv} for the scalar and vector Green's function and their mass derivatives in the general gauge Green's function~\eqref{tensor_propagator}, using also the relations~\eqref{hadamard_4d_vector_hatv_in_v}, \eqref{hadamard_4d_hatv_bdy}, \eqref{hadamard_4d_v_massexpansion} and~\eqref{hadamard_4d_tensor_uv_bdy} as well as the definitions of $\mathfrak{m}^2$~\eqref{mh_def} and $\mathbf{M}^2$~\eqref{mzeta_def}. After a straightforward but lengthy calculation, it follows that
\begin{equation}
\label{hadamard_4d_tensor_xizeta}
G^{\xi,\zeta}_{\mu\nu\alpha'\beta'} = - \frac{\mathi}{8 \pi^2} \left[ \frac{U^{\xi,\zeta}_{\mu\nu\alpha'\beta'}}{\sigma} + V^{\xi,\zeta}_{\mu\nu\alpha'\beta'} \ln \left( \mu^2 \sigma \right) + W^{\xi,\zeta}_{\mu\nu\alpha'\beta'} \right] \eqend{,}
\end{equation}
where similar to the vector case the expansion of $U^{\xi,\zeta}$ now contains negative powers of $\sigma$, namely
\begin{equation}
U^{\xi,\zeta}_{\mu\nu\alpha'\beta'} = \sum_{k=-3}^0 U^{\xi,\zeta(k)}_{\mu\nu\alpha'\beta'} \sigma^k \eqend{,}
\end{equation}
while $V^{\xi,\zeta}$ and $W^{\xi,\zeta}$ have the usual asymptotic expansion
\begin{subequations}\begin{align}
V^{\xi,\zeta}_{\mu\nu\alpha'\beta'} &= \sum_{k=0}^\infty V^{\xi,\zeta(k)}_{\mu\nu\alpha'\beta'} \sigma^k \eqend{,} \\
W^{\xi,\zeta}_{\mu\nu\alpha'\beta'} &= \sum_{k=0}^\infty W^{\xi,\zeta(k)}_{\mu\nu\alpha'\beta'} \sigma^k \eqend{.}
\end{align}\end{subequations}
We obtain
\begin{equation}
\label{hadamard_4d_tensor_um3}
U^{\xi,\zeta(-3)}_{\mu\nu\alpha'\beta'} = 0 \eqend{,}
\end{equation}
\begin{equation}
U^{\xi,\zeta(-2)}_{\mu\nu\alpha'\beta'} = - 2 c(\xi,\zeta) \sigma_\mu \sigma_\nu \sigma_{\alpha'} \sigma_{\beta'} \sqrt{\Delta} \eqend{,}
\end{equation}
\begin{widetext}
\begin{equation}
U^{\xi,\zeta(-1)}_{\mu\nu\alpha'\beta'} = \left[ 2 (\xi-1) \sigma_{(\mu} g_{\nu)(\alpha'} \sigma_{\beta')} - \frac{(1-\zeta)}{(1-2\zeta)} (g\sdot\sigma^2)_{\mu\nu\alpha'\beta'} + c(\xi,\zeta) (\sigma^3\sdot\nabla)_{\mu\nu\alpha'\beta'} \right] \sqrt{\Delta} - 2 \sigma_\mu \sigma_\nu \sigma_{\alpha'} \sigma_{\beta'} \Delta V^{\xi,\zeta(2)} \eqend{,}
\end{equation}
\begin{equation}\begin{split}
U^{\xi,\zeta(0)}_{\mu\nu\alpha'\beta'} &= \left( g_{\alpha'(\mu} g_{\nu)\beta'} - \frac{1}{2} g_{\mu\nu} g_{\alpha'\beta'} \right) \sqrt{\Delta} - (\xi-1) \left( \sigma_{\alpha'(\mu} g_{\nu)\beta'} + \sigma_{\beta'(\mu} g_{\nu)\alpha'} \right) \sqrt{\Delta} - 2 (\xi-1) \sigma_{(\mu} V^{\mathfrak{m}^2,1(0)}_{\nu)(\alpha'} \sigma_{\beta')} \\
&\quad- (\xi-1) \left[ \sigma_\mu \nabla_{(\alpha'} \left( \sqrt{\Delta} \, g_{\beta')\nu} \right) + \sigma_\nu \nabla_{(\alpha'} \left( \sqrt{\Delta} \, g_{\beta')\mu} \right) + \sigma_{\alpha'} \nabla_{(\mu} \left( \sqrt{\Delta} \, g_{\nu)\beta'} \right) + \sigma_{\beta'} \nabla_{(\mu} \left( \sqrt{\Delta} \, g_{\nu)\alpha'} \right) \right] \\
&\quad+ \frac{(1-\zeta)}{(1-2\zeta)} (g\sdot\sigma\sdot\nabla)_{\mu\nu\alpha'\beta'} \sqrt{\Delta} - c(\xi,\zeta) (\sigma^2\sdot\nabla^2)_{\mu\nu\alpha'\beta'} \sqrt{\Delta} + \frac{(1-\zeta)}{(1-2\zeta)} (g\sdot\sigma^2)_{\mu\nu\alpha'\beta'} \left[ V_0^{(0)} + \frac{(1-3\zeta)}{4 (1-2\zeta)} \mathfrak{m}^2 \sqrt{\Delta} \right] \\
&\quad+ 2 (\sigma^3\sdot\nabla)_{\mu\nu\alpha'\beta'} \Delta V^{\xi,\zeta(2)} + 6 \sigma_\mu \sigma_\nu \sigma_{\alpha'} \sigma_{\beta'} \Delta V^{\xi,\zeta(3)} \eqend{,}
\end{split}\end{equation}
\begin{equation}\begin{split}
V^{\xi,\zeta(0)}_{\mu\nu\alpha'\beta'} &= V^{1,1(0)}_{\mu\nu\alpha'\beta'} - (\xi-1) \left[ \nabla_\mu \nabla_{(\alpha'} \left( \sqrt{\Delta} \, g_{\beta')\nu} \right) + \nabla_\nu \nabla_{(\alpha'} \left( \sqrt{\Delta} \, g_{\beta')\mu} \right) \right] - 2 (\xi-1) \sigma_{(\mu} V^{\mathfrak{m}^2,1(1)}_{\nu)(\alpha'} \sigma_{\beta')} \\
&\quad- (\xi-1) \left[ \sigma_{\alpha'(\mu} V^{\mathfrak{m}^2,1(0)}_{\nu)\beta'} + \sigma_{\beta'(\mu} V^{\mathfrak{m}^2,1(0)}_{\nu)\alpha'} + \sigma_\mu \nabla_{(\alpha'} V^{\mathfrak{m}^2,1(0)}_{|\nu|\beta')} + \sigma_\nu \nabla_{(\alpha'} V^{\mathfrak{m}^2,1(0)}_{|\mu|\beta')} + 2 \nabla_{(\mu} V^{\mathfrak{m}^2,1(0)}_{\nu)(\alpha'} \sigma_{\beta')} \right] \\
&\quad+ \frac{(1-\zeta)}{(1-2\zeta)} (g\sdot\sigma^2)_{\mu\nu\alpha'\beta'} V_0^{(1)} + \frac{(1-\zeta)}{(1-2\zeta)} \left[ (g\sdot\sigma\sdot\nabla)_{\mu\nu\alpha'\beta'} + \frac{(1-3\zeta)}{4 (1-2\zeta)} \mathfrak{m}^2 (g\sdot\sigma^2)_{\mu\nu\alpha'\beta'} \right] V_0^{(0)} \\
&\quad+ \frac{(1-\zeta)}{(1-2\zeta)} \left[ g_{\alpha'\beta'} \nabla_\mu \nabla_\nu + g_{\mu\nu} \nabla_{\alpha'} \nabla_{\beta'} + \frac{(1-3\zeta)}{4 (1-2\zeta)} \mathfrak{m}^2 (g\sdot\sigma\sdot\nabla)_{\mu\nu\alpha'\beta'} + \frac{(1-5\zeta+7\zeta^2)}{24 (1-2\zeta)^2} \mathfrak{m}^4 (g\sdot\sigma^2)_{\mu\nu\alpha'\beta'} \right] \sqrt{\Delta} \\
&\quad- c(\xi,\zeta) (\sigma\sdot\nabla^3)_{\mu\nu\alpha'\beta'} \sqrt{\Delta} + 2 (\sigma^2\sdot\nabla^2)_{\mu\nu\alpha'\beta'} \Delta V^{\xi,\zeta(2)} + 6 (\sigma^3\sdot\nabla)_{\mu\nu\alpha'\beta'} \Delta V^{\xi,\zeta(3)} + 24 \sigma_\mu \sigma_\nu \sigma_{\alpha'} \sigma_{\beta'} \Delta V^{\xi,\zeta(4)}
\end{split}\end{equation}
and (for $k \geq 1$)
\begin{equation}\begin{split}
V^{\xi,\zeta(k)}_{\mu\nu\alpha'\beta'} &= V^{1,1(k)}_{\mu\nu\alpha'\beta'} - \frac{\xi-1}{k} \left( \nabla_\mu \nabla_{(\alpha'} V^{\mathfrak{m}^2,1(k-1)}_{|\nu|\beta')} + \nabla_\nu \nabla_{(\alpha'} V^{\mathfrak{m}^2,1(k-1)}_{|\mu|\beta')} \right) - 2 (k+1) (\xi-1) \sigma_{(\mu} V^{\mathfrak{m}^2,1(k+1)}_{\nu)(\alpha'} \sigma_{\beta')} \\
&\quad- (\xi-1) \left[ \sigma_{\alpha'(\mu} V^{\mathfrak{m}^2,1(k)}_{\nu)\beta'} + \sigma_{\beta'(\mu} V^{\mathfrak{m}^2,1(k)}_{\nu)\alpha'} + \sigma_\mu \nabla_{(\alpha'} V^{\mathfrak{m}^2,1(k)}_{|\nu|\beta')} + \sigma_\nu \nabla_{(\alpha'} V^{\mathfrak{m}^2,1(k)}_{|\mu|\beta')} + 2 \nabla_{(\mu} V^{\mathfrak{m}^2,1(k)}_{\nu)(\alpha'} \sigma_{\beta')} \right] \\
&\quad+ \frac{2 (1-\zeta)}{(1-2\zeta)} \left[ (k+2) (k+1) (g\sdot\sigma^2)_{\mu\nu\alpha'\beta'} \frac{V^{(k+2)}_{\mathbf{M}^2} - V^{(k+2)}_{\mathfrak{m}^2}}{\mathbf{M}^2 - \mathfrak{m}^2} + (k+1) (g\sdot\sigma\sdot\nabla)_{\mu\nu\alpha'\beta'} \frac{V^{(k+1)}_{\mathbf{M}^2} - V^{(k+1)}_{\mathfrak{m}^2}}{\mathbf{M}^2 - \mathfrak{m}^2} \right] \\
&\quad+ \frac{2 (1-\zeta)}{(1-2\zeta)} \left[ \left( g_{\alpha'\beta'} \nabla_\mu \nabla_\nu + g_{\mu\nu} \nabla_{\alpha'} \nabla_{\beta'} \right) \frac{V^{(k)}_{\mathbf{M}^2} - V^{(k)}_{\mathfrak{m}^2}}{\mathbf{M}^2 - \mathfrak{m}^2} \right] \\
&\quad+ \nabla_\mu \nabla_\nu \nabla_{\alpha'} \nabla_{\beta'} \Delta V^{\xi,\zeta(k)} + (k+1) (\sigma\sdot\nabla^3)_{\mu\nu\alpha'\beta'} \Delta V^{\xi,\zeta(k+1)} + (k+2) (k+1) (\sigma^2\sdot\nabla^2)_{\mu\nu\alpha'\beta'} \Delta V^{\xi,\zeta(k+2)} \\
&\quad+ (k+3) (k+2) (k+1) (\sigma^3\sdot\nabla)_{\mu\nu\alpha'\beta'} \Delta V^{\xi,\zeta(k+3)} + (k+4) (k+3) (k+2) (k+1) \sigma_\mu \sigma_\nu \sigma_{\alpha'} \sigma_{\beta'} \Delta V^{\xi,\zeta(k+4)} \eqend{,}
\end{split}\end{equation}
with the abbreviations
\begin{subequations}\label{tensor_abbr}\begin{align}
\Delta V^{\xi,\zeta(k)} &\equiv - (\xi-1) \frac{4 (1-\zeta)}{(1-2\zeta)} \frac{\hat{V}^{(k)}_{\mathbf{M}^2} - \hat{V}^{(k)}_{\mathfrak{m}^2}}{\mathbf{M}^2 - \mathfrak{m}^2} + \frac{4 (1-\zeta)^2 (3-\xi)}{(1-2\zeta)^2} \frac{V^{(k)}_{\mathbf{M}^2} - V^{(k)}_{\mathfrak{m}^2} - \left( \mathbf{M}^2 - \mathfrak{m}^2 \right) \hat{V}^{(k)}_{\mathbf{M}^2}}{\left( \mathbf{M}^2 - \mathfrak{m}^2 \right)^2} \eqend{,} \\
(g\sdot\sigma^2)_{\mu\nu\alpha'\beta'} &\equiv g_{\alpha'\beta'} \sigma_\mu \sigma_\nu + g_{\mu\nu} \sigma_{\alpha'} \sigma_{\beta'} \eqend{,} \\
(\sigma^3\sdot\nabla)_{\mu\nu\alpha'\beta'} &\equiv 2 \sigma_\mu \sigma_\nu \sigma_{(\alpha'} \nabla_{\beta')} + 2 \sigma_{\alpha'} \sigma_{\beta'} \sigma_{(\mu} \nabla_{\nu)} + 4 \sigma_{(\mu} \sigma_{\nu)(\alpha'} \sigma_{\beta')} + \sigma_\mu \sigma_\nu \sigma_{\alpha'\beta'} + \sigma_{\mu\nu} \sigma_{\alpha'} \sigma_{\beta'} \eqend{,} \\
(g\sdot\sigma\sdot\nabla)_{\mu\nu\alpha'\beta'} &\equiv 2 g_{\alpha'\beta'} \sigma_{(\mu} \nabla_{\nu)} + 2 g_{\mu\nu} \sigma_{(\alpha'} \nabla_{\beta')} + g_{\alpha'\beta'} \sigma_{\mu\nu} + g_{\mu\nu} \sigma_{\alpha'\beta'} \eqend{,} \\
\begin{split}
(\sigma^2\sdot\nabla^2)_{\mu\nu\alpha'\beta'} &\equiv \sigma_{\alpha'} \sigma_{\beta'} \nabla_\mu \nabla_\nu + \sigma_\mu \sigma_\nu \nabla_{\alpha'} \nabla_{\beta'} + 2 \sigma_\mu \sigma_{(\alpha'} \nabla_{\beta')} \nabla_\nu + 2 \sigma_\nu \sigma_{(\alpha'} \nabla_{\beta')} \nabla_\mu + 2 \sigma_{\mu\nu} \sigma_{(\alpha'} \nabla_{\beta')} + 2 \sigma_{\alpha'\beta'} \sigma_{(\mu} \nabla_{\nu)} \\
&\quad+ 4 \sigma_{(\alpha'} \sigma_{\beta')(\mu} \nabla_{\nu)} + 4 \sigma_{(\mu} \sigma_{\nu)(\alpha'} \nabla_{\beta')} + \sigma_{\mu\nu} \sigma_{\alpha'\beta'} + 2 \sigma_{\mu(\alpha'} \sigma_{\beta')\nu} + 2 \sigma_{(\mu} \sigma_{\nu)\alpha'\beta'} + 2 \sigma_{(\alpha'} \sigma_{\beta')\mu\nu} \eqend{,}
\end{split} \\
\begin{split}
(\sigma\sdot\nabla^3)_{\mu\nu\alpha'\beta'} &\equiv 2 \sigma_{(\mu} \nabla_{\nu)} \nabla_{\alpha'} \nabla_{\beta'} + 2 \sigma_{(\alpha'} \nabla_{\beta')} \nabla_\mu \nabla_\nu + 2 \sigma_{\mu(\alpha'} \nabla_{\beta')} \nabla_\nu + 2 \sigma_{\nu(\alpha'} \nabla_{\beta')} \nabla_\mu \\
&\quad+ \sigma_{\mu\nu} \nabla_{\alpha'} \nabla_{\beta'} + \sigma_{\alpha'\beta'} \nabla_\mu \nabla_\nu + 2 \sigma_{\alpha'\beta'(\mu} \nabla_{\nu)} + 2 \sigma_{\mu\nu(\alpha'} \nabla_{\beta')} + \sigma_{\mu\nu\alpha'\beta'} \eqend{,}
\end{split} \\
c(\xi,\zeta) &\equiv \frac{(1-\zeta) [ (\xi-1) (1-3\zeta) + 2 (1-\zeta) ]}{2 (1-2\zeta)^2} \eqend{.}
\end{align}\end{subequations}
\end{widetext}
Again, we have delegated the even lengthier expressions for the $W$ coefficients to Appendix~\ref{appendix_w}. Similarly to the vector case, the general-gauge graviton Green's function~\eqref{tensor_propagator} is less singular than one would naively expect. Since the coefficient $U^{(0)}_{m^2}$ of the most singular term in the Hadamard expansion of the scalar Green's function~\eqref{hadamard_4d_scalar} is mass-independent, the term proportional to $\sigma^{-1}$ disappears both from the difference of scalar Green's functions and their mass derivative in Eq.~\eqref{tensor_propagator}, leaving the logarithmic term $\ln( \mu^2 \sigma )$ as the most singular. Because four derivatives act on it, one would expect a term proportional to $\sigma^{-4}$ in the Hadamard expansion~\eqref{hadamard_4d_tensor_xizeta} for a general gauge. Nevertheless, the coefficient of this term turns out to vanish~\eqref{hadamard_4d_tensor_um3}, and the most singular term is proportional to $\sigma^{-3}$. So far, the reason for this cancellation is unclear.

\subsection{Generalization to \texorpdfstring{$n$}{n} dimensions}

Since already in four dimensions the formulas for the Hadamard expansions in general gauges (in particular the state-dependent $W$ coefficients) become quite complicated, we only indicate how one proceeds in $n \neq 4$ dimensions. To actually calculate the coefficients themselves, and in addition their covariant expansion that is needed to calculate the expectation values of composite operators, the use of a computer algebra system is highly recommended~\cite{ottewillwardell2011}.

\subsubsection*{Scalar field}

In $n$ dimensions, the Hadamard expansion for a scalar field has the form (see, e.g., Ref.~\cite{decaninifolacci2008})
\begin{equation}
G_{m^2} = - \mathi c_n \left[ \frac{U_{m^2}}{\sigma^{n/2-1}} + V_{m^2} \ln \left( \mu^2 \sigma \right) + W_{m^2} \right] \eqend{,}
\end{equation}
where the constant $c_n$ is given by
\begin{equation}
c_n = \begin{cases} 1/(4\pi) & n = 2 \\ \dfrac{\Gamma\left( \frac{n}{2} - 1 \right)}{2 (2\pi)^\frac{n}{2}} & n > 2 \eqend{,} \end{cases}
\end{equation}
and the asymptotic expansion of the biscalars $\{ U/V/W \}_{m^2}$ is of the form
\begin{equation}
\{ U/V/W \}_{m^2} = \sum_{k=0}^\infty \{ U/V/W \}_{m^2}^{(k)} \sigma^k \eqend{,}
\end{equation}
where $V_{m^2}^{(k)} = 0$ in odd dimensions and $U_{m^2}^{(k)} = 0$ for $k > n/2-2$ in even dimensions. That is, in odd dimensions the logarithmic term is absent, in $n = 2$ dimensions $U_{m^2} = 0$, and in even dimensions greater than $n = 4$ there are terms more singular than $\sigma^{-1}$. The recursion relations~\eqref{hadamard_4d_scalar_recursion} now read
\begin{subequations}\begin{align}
Q_{2k+2} U_{m^2}^{(k+1)} &= - \frac{2}{2k+4-n} P_{m^2} U_{m^2}^{(k)} \eqend{,} \\
Q_{2k+n} V_{m^2}^{(k+1)} &= - \frac{1}{k+1} P_{m^2} V_{m^2}^{(k)} \eqend{,} \\
Q_{2k+n} W_{m^2}^{(k+1)} &= - \frac{1}{k+1} \left( P_{m^2} W_{m^2}^{(k)} + Q_{4k+2+n} V_{m^2}^{(k+1)} \right)
\end{align}\end{subequations}
with the Klein--Gordon operator $P_{m^2}$~\eqref{scalar_p_def}, and must be solved with the boundary conditions
\begin{equation}
U_{m^2}^{(0)} = 0 \eqend{,} \qquad V_{m^2}^{(0)} = - \sqrt{\Delta}
\end{equation}
in $n = 2$ dimensions,
\begin{equation}
U_{m^2}^{(0)} = \sqrt{\Delta} \eqend{,} \qquad Q_{n-2} V_{m^2}^{(0)} = - P_{m^2} U_{m^2}^{(n/2-2)}
\end{equation}
in even dimensions greater than 2, and
\begin{equation}
U_{m^2}^{(0)} = \sqrt{\Delta} \eqend{,} \qquad V_{m^2}^{(0)} = 0
\end{equation}
in odd dimensions, while $W_{m^2}^{(0)}$ is always a solution of the Klein--Gordon equation $P_{m^2} W_{m^2}^{(0)} = 0$. Formulas for the mass derivatives can then be derived in exactly the same way as for $n = 4$.

\subsubsection*{Vector field}

In Feynman gauge $\xi = 1$, we have the direct generalization of the scalar formula for the propagator:
\begin{equation}
G^{m^2,1}_{\nu\beta'} = - \mathi c_n \left[ \frac{U^{m^2,1}_{\nu\beta'}}{\sigma^{n/2-1}} + V^{m^2,1}_{\nu\beta'} \ln \left( \mu^2 \sigma \right) + W^{m^2,1}_{\nu\beta'} \right] \eqend{,}
\end{equation}
with the asymptotic expansions
\begin{equation}
\{ U/V/W \}^{m^2,1}_{\nu\beta'} = \sum_{k=0}^\infty \{ U/V/W \}^{m^2,1(k)}_{\nu\beta'} \sigma^k \eqend{,}
\end{equation}
where $V^{m^2,1(k)}_{\nu\beta'} = 0$ in odd dimensions and $U^{m^2,1(k)}_{\nu\beta'} = 0$ for $k > n/2-2$ in even dimensions. The recursion relations are
\begin{subequations}\begin{align}
Q_{2k+2} U^{m^2,1(k+1)}_{\nu\beta'} &= - \frac{2}{2k+4-n} g_{\nu\mu} P_{m^2,1}^{\mu\rho} U^{m^2,1(k)}_{\rho\beta'} \eqend{,} \\
Q_{2k+n} V^{m^2,1(k+1)}_{\nu\beta'} &= - \frac{1}{k+1} g_{\nu\mu} P_{m^2,1}^{\mu\rho} V^{m^2,1(k)}_{\rho\beta'} \eqend{,} \\
\begin{split}
Q_{2k+n} W^{m^2,1(k+1)}_{\nu\beta'} &= - \frac{1}{k+1} g_{\nu\mu} P_{m^2,1}^{\mu\rho} W^{m^2,1(k)}_{\rho\beta'} \\
&\quad- \frac{1}{k+1} Q_{4k+2+n} V^{m^2,1(k+1)}_{\nu\beta'}
\end{split}
\end{align}\end{subequations}
with the operator $P_{m^2,1}^{\mu\rho}$ defined in Eq.~\eqref{vector_p_def}, and are solved with the boundary conditions
\begin{equation}
U^{m^2,1(0)}_{\nu\beta'} = 0 \eqend{,} \qquad V^{m^2,1(0)}_{\nu\beta'} = - \sqrt{\Delta} \, g_{\nu\beta'}
\end{equation}
in $n = 2$ dimensions,
\begin{subequations}\begin{align}
U^{m^2,1(0)}_{\nu\beta'} &= \sqrt{\Delta} \, g_{\nu\beta'} \eqend{,} \\
Q_{n-2} V^{m^2,1(0)}_{\nu\beta'} &= - g_{\nu\mu} P_{m^2,1}^{\mu\rho} U^{m^2,1(n/2-2)}_{\rho\beta'}
\end{align}\end{subequations}
in even dimensions greater than 2, and
\begin{equation}
U^{m^2,1(0)}_{\nu\beta'} = \sqrt{\Delta} \, g_{\nu\beta'} \eqend{,} \qquad V^{m^2,1(0)}_{\nu\beta'} = 0
\end{equation}
in odd dimensions, and $W^{m^2,1(0)}_{\nu\beta'}$ being a solution to $P_{m^2,1}^{\mu\rho} W^{m^2,1(0)}_{\rho\beta'} = 0$.

In a general gauge, the Hadamard expansion is again obtained by inserting the scalar and Feynman gauge vector Hadamard expansions into Eq.~\eqref{vector_propagator}. We refrain from giving the general expression here, and only note that the most singular term is of order $\sigma^{-n/2}$.

\subsubsection*{Tensor field}

In the gauge $\xi = \zeta = 1$, we again have the direct generalization of the scalar formula for the propagator:
\begin{equation}
G^{1,1}_{\mu\nu\alpha'\beta'} = - \mathi c_n \left[ \frac{U^{1,1}_{\mu\nu\alpha'\beta'}}{\sigma^{n/2-1}} + V^{1,1}_{\mu\nu\alpha'\beta'} \ln \left( \mu^2 \sigma \right) + W^{1,1}_{\mu\nu\alpha'\beta'} \right] \eqend{,}
\end{equation}
with the asymptotic expansions
\begin{equation}
\{ U/V/W \}^{1,1}_{\mu\nu\alpha'\beta'} = \sum_{k=0}^\infty \{ U/V/W \}^{1,1(k)}_{\mu\nu\alpha'\beta'} \sigma^k \eqend{,}
\end{equation}
where $V^{1,1(k)}_{\mu\nu\alpha'\beta'} = 0$ in odd dimensions and $U^{1,1(k)}_{\mu\nu\alpha'\beta'} = 0$ for $k > n/2-2$ in even dimensions. The recursion relations are
\begin{subequations}\begin{align}
Q_{2k+2} U^{1,1(k+1)}_{\mu\nu\alpha'\beta'} &= - \frac{2}{2k+4-n} g_{\mu\rho} g_{\nu\sigma} P_{1,1}^{\rho\sigma\gamma\delta} U^{1,1(k)}_{\gamma\delta\alpha'\beta'} \eqend{,} \\
Q_{2k+n} V^{1,1(k+1)}_{\mu\nu\alpha'\beta'} &= - \frac{1}{k+1} g_{\mu\rho} g_{\nu\sigma} P_{1,1}^{\rho\sigma\gamma\delta} V^{1,1(k)}_{\gamma\delta\alpha'\beta'} \eqend{,} \\
\begin{split}
Q_{2k+n} W^{1,1(k+1)}_{\mu\nu\alpha'\beta'} &= - \frac{1}{k+1} g_{\mu\rho} g_{\nu\sigma} P_{1,1}^{\rho\sigma\gamma\delta} W^{1,1(k)}_{\gamma\delta\alpha'\beta'} \\
&\quad- \frac{1}{k+1} Q_{4k+2+n} V^{1,1(k+1)}_{\mu\nu\alpha'\beta'}
\end{split}
\end{align}\end{subequations}
with the operator $P_{1,1}^{\mu\nu\rho\sigma}$ defined in Eq.~\eqref{tensor_p_def} and the boundary conditions
\begin{equation}
U^{1,1(0)}_{\mu\nu\alpha'\beta'} = \sqrt{\Delta} \left( g_{\alpha'(\mu} g_{\nu)\beta'} - \frac{1}{n-2} g_{\mu\nu} g_{\alpha'\beta'} \right) \eqend{,}
\end{equation}
in $n > 2$ dimensions,
\begin{equation}
Q_{n-2} V^{1,1(0)}_{\mu\nu\alpha'\beta'} = - g_{\mu\rho} g_{\nu\sigma} P_{1,1}^{\rho\sigma\gamma\delta} U^{1,1(n/2-2)}_{\gamma\delta\alpha'\beta'}
\end{equation}
in even dimensions greater than 2, and
\begin{equation}
V^{1,1(0)}_{\mu\nu\alpha'\beta'} = 0
\end{equation}
in odd dimensions, and $W^{1,1(0)}_{\mu\nu\alpha'\beta'}$ is a solution to $P_{1,1}^{\rho\sigma\gamma\delta} W^{1,1(0)}_{\gamma\delta\alpha'\beta'} = 0$.

In a general gauge, the Hadamard expansion is again obtained by inserting the scalar and vector Hadamard expansions into Eq.~\eqref{tensor_propagator}. The corresponding expressions are extremely lengthy, and we note that while naively the most singular term is of order $\sigma^{-n/2-2}$, it is again possible that as in four dimensions its coefficient may vanish.

\section{Outlook}
\label{section_outlook}

We have studied vector and tensor Green's functions in different linear covariant gauges, derived divergence and trace identities and calculated their Hadamard expansions. Although the classical gauge theories are clearly independent of the choice of gauge fixing, the issue of gauge-fixing independence at the quantum level is technically much more involved (see, e.g., Ref.~\cite{allenottewill1992} for a proof of the independence of the stress tensor of the gauge parameter for electrodynamics). A suitable formalism to study these issues is BRST quantization, where classical observables are invariant under the action of the (classical) nilpotent BRST differential $\brst$, which generalizes the gauge symmetry to the additional fields introduced in the BRST formalism as explained in Sec.~\ref{section_ward}. Furthermore, two observables are identified if they differ by an $\brst$-exact term, such that one needs to study the cohomology of $\brst$. At the quantum level, the BRST differential needs to be extended to a quantum BRST differential $\mathsf{q}$~\cite{froebhollandhollands2016,taslimitehrani2017}, which differs from $\brst$ by corrections of order $\hbar$ (and higher), and the (renormalized) observables are in the cohomology of $\mathsf{q}$. Formally, the independence of the correlation functions of these observables from the choice of gauge fixing follows from the BRST invariance of the full action $S$ including counterterms $\brst S = 0$, in a regularization scheme where $\mathsf{q} = \brst$ (such as dimensional regularization). To prove this independence rigorously, which can be done in the framework of algebraic quantum field theory on curved spacetimes, it is of course necessary to first construct the theory for different choices of the gauge fixing. To construct the algebra of (composite) field operators in the free theory, one needs to know the Hadamard parametrix in order to define local and covariant Wick powers, and to construct the interacting algebra one needs to know the retarded Green's functions, both of which we provide in this work for the class of linear covariant gauges. It then turns out that the cohomologies of $\mathsf{q}$ for two different gauges are isomorphic~\cite{taslimitehrani}, i.e., there is a one-to-one map between observables calculated in two different gauges.

\begin{acknowledgments}
We thank Chris Fewster, Thomas-Paul Hack, Stefan Hollands and Jochen Zahn for discussions, and Stefan Hollands for originally suggesting the problem and for a critical reading of the manuscript. This work is part of a project that has received funding from the European Union's Horizon 2020 research and innovation programme under the Marie Sk{\l}odowska-Curie grant agreement No. 702750 ``QLO-QG''. M.T.T.~thanks the Max-Planck-Institut f{\"u}r Mathematik in den Naturwissenschaften and its International Max Planck Research School (IMPRS) for financial support.
\end{acknowledgments}

\appendix

\section{Hadamard coefficients in Riemann normal coordinates}
\label{appendix_riemann}

The explicit solution to the recursion relations~\eqref{hadamard_4d_scalar_recursion} can be given in Riemann normal coordinates. For this, we first rewrite the differential operator $Q_k$~\eqref{qk_def} in the form
\begin{equation}
Q_k F = \sqrt{\Delta} \Big( 2 \nabla^\mu \sigma \nabla_\mu + k \Big) \left( \frac{F}{\sqrt{\Delta}} \right) \eqend{.}
\end{equation}
Riemann normal coordinates are such that the geodesics from $x'$ to $x$ are straight lines:
\begin{equation}
y^\mu(\lambda) = (x')^\mu + \lambda \left[ x^\mu - (x')^\mu \right] \eqend{.}
\end{equation}
It follows that $\sigma$ is given by
\begin{equation}
\sigma(x,x') = \frac{1}{2} g_{\mu\nu}(x) \frac{\partial y^\mu}{\partial \lambda} \frac{\partial y^\nu}{\partial \lambda} \eqend{,}
\end{equation}
and we calculate
\begin{equation}\begin{split}
&\left( 2 \nabla^\mu \sigma \nabla_\mu + k \right) F(y,x') = \left( 2 \frac{\partial y^\nu}{\partial \lambda} \nabla_\nu + k \right) F(y,x') \\
&\quad= \left( 2 \lambda \partial_\lambda + k \right) F(y,x') = 2 \lambda^{-\frac{k-2}{2}} \partial_\lambda \left[ \lambda^\frac{k}{2} F(y,x') \right] \eqend{.}
\end{split}\end{equation}
We thus obtain that the unique smooth solution of $Q_k F = J$ is given for all $k > 0$ by
\begin{equation}
\label{qk_smooth_solution}
\left[ \frac{F}{\sqrt{\Delta}} \right](x,x') = \frac{1}{2} \int_0^1 \left[ \frac{J}{\sqrt{\Delta}} \right](y,x') \lambda^\frac{k-2}{2} \total \lambda \eqend{.}
\end{equation}
The recursion relations for the scalar Hadamard coefficients~\eqref{hadamard_4d_scalar_recursion} can now be explicitly solved and read
\begin{subequations}\begin{align}
&V_{m^2}^{(k+1)} = - \frac{\sqrt{\Delta}}{2(k+1)} \int_0^1 \left[ \frac{P_{m^2} V_{m^2}^{(k)}}{\sqrt{\Delta}} \right](y,x') \lambda^{k+1} \total \lambda \eqend{,} \\
\begin{split}
&W_{m^2}^{(k+1)} = - \frac{1}{k+1} V_{m^2}^{(k+1)} - \frac{\sqrt{\Delta}}{2(k+1)} \\
&\quad\times \int_0^1 \left[ \frac{P_{m^2} W_{m^2}^{(k)} + 2 (k+1) V_{m^2}^{(k+1)}}{\sqrt{\Delta}} \right](y,x') \lambda^k \total \lambda \eqend{,}
\end{split}
\end{align}\end{subequations}
and the boundary condition~\eqref{hadamard_4d_v0_bdy} for $V_{m^2}^{(0)}$ can be written as
\begin{equation}
\label{hadamard_4d_v_bdy}
V_{m^2}^{(0)} = - \frac{1}{2} \sqrt{\Delta} \int_0^1 \left[ \frac{P_{m^2} \sqrt{\Delta}}{\sqrt{\Delta}} \right](y,x') \total \lambda \eqend{.}
\end{equation}
For the vector and tensor coefficients we obtain similar expressions.

\section{Formulas for the \texorpdfstring{$W$}{W} coefficients}
\label{appendix_w}

Assuming that the Feynman propagators or Wightman functions in two different gauges are related in the same way as the Green's functions~\eqref{vector_propagator}, \eqref{vector_propagator_massless} and~\eqref{tensor_propagator}, we can also determine the relation between the $W$ coefficients in the same way as for the $U$ and $V$ coefficients. For the vector, this gives
\begin{equation}\begin{split}
&W^{m^2,\xi(k)}_{\nu\beta'} = W^{m^2,1(k)}_{\nu\beta'} - \nabla_\nu \nabla_{\beta'} \Delta W_{\xi,m^2}^{(k)} \\
&\quad- (\sigma\sdot\nabla)_{\nu\beta'} \left[ (k+1) \Delta W_{\xi,m^2}^{(k+1)} + \Delta V_{\xi,m^2}^{(k+1)} \right] \\
&\quad- \sigma_\nu \sigma_{\beta'} \left[ (k+2) (k+1) \Delta W_{\xi,m^2}^{(k+2)} + (2k+3) \Delta V_{\xi,m^2}^{(k+2)} \right] \eqend{,}
\end{split}\end{equation}
where we defined
\begin{equation}
\Delta W_{\xi,m^2}^{(k)} \equiv \frac{W_{\xi m^2}^{(k)} - W_{m^2}^{(k)}}{m^2} \eqend{,} \quad \Delta V_{\xi,m^2}^{(k)} \equiv \frac{V_{\xi m^2}^{(k)} - V_{m^2}^{(k)}}{m^2} \eqend{,}
\end{equation}
and
\begin{equation}
(\sigma\sdot\nabla)_{\nu\beta'} \equiv 2 \sigma_{(\nu} \nabla_{\beta')} + \sigma_{\nu\beta'} \eqend{,}
\end{equation}
and in the massless limit we have [using the relation~\eqref{hadamard_4d_hatv_in_v}]
\begin{equation}
\Delta W_{\xi,m^2}^{(k)} \to (\xi-1) \hat{W}_0^{(k)} \eqend{,} \qquad \Delta V_{\xi,m^2}^{(k)} \to \frac{\xi-1}{2k} V_0^{(k-1)} \eqend{.}
\end{equation}

For the Hadamard expansion coefficients of the mass derivative of the vector Green's function~\eqref{hadamard_4d_vector_xi_massderiv}, we obtain
\begin{subequations}\label{hadamard_4d_vector_uv_massexpansion}\begin{align}
&\hat{U}^{m^2,\xi(0)}_{\nu\beta'} = - \frac{\xi-1}{4} \sqrt{\Delta} \sigma_\nu \sigma_{\beta'} \eqend{,} \\
\begin{split}
&\hat{V}^{m^2,\xi(0)}_{\nu\beta'} = \frac{1}{2} \sqrt{\Delta} \, g_{\nu\beta'} - \frac{\xi^2-1}{8} (\sigma\sdot\nabla)_{\nu\beta'} \sqrt{\Delta} \\
&\quad- \frac{1}{8} \left[ (\xi^2-1) V_0^{(0)} + \frac{\xi^3-1}{6} m^2 \sqrt{\Delta} \right] \sigma_\nu \sigma_{\beta'} \eqend{,}
\end{split} \\
\begin{split}
&\hat{V}^{m^2,\xi(k)}_{\nu\beta'} = \frac{1}{2 k} V^{m^2,1(k-1)}_{\nu\beta'} - \frac{1}{2k} \nabla_\nu \nabla_{\beta'} \delta V_{\xi, m^2}^{(k)} \\
&\quad- \frac{1}{2} (\sigma\sdot\nabla)_{\nu\beta'} \delta V_{\xi, m^2}^{(k+1)} - \frac{k+1}{2} \sigma_\nu \sigma_{\beta'} \delta V_{\xi, m^2}^{(k+2)} \quad (k \geq 1) \eqend{,}
\end{split} \\
\begin{split}
&\hat{W}^{m^2,\xi(k)}_{\nu\beta'} = \hat{W}^{m^2,1(k)}_{\nu\beta'} - \frac{1}{2k} \nabla_\nu \nabla_{\beta'} \delta W_{\xi, m^2}^{(k)} \\
&\quad- \frac{1}{2 (k+1)} (\sigma\sdot\nabla)_{\nu\beta'} \left[ (k+1) \delta W_{\xi, m^2}^{(k+1)} + \delta V_{\xi, m^2}^{(k+1)} \right] \\
&\quad- \frac{\sigma_\nu \sigma_{\beta'}}{2 (k+2)} \left[ (k+1) (k+2) \delta W_{\xi, m^2}^{(k+2)} + (2k+3) \delta V_{\xi, m^2}^{(k+2)} \right] \eqend{,}
\end{split}
\end{align}\end{subequations}
where we defined
\begin{subequations}\begin{align}
\begin{split}
\delta V_{\xi, m^2}^{(k+1)} &\equiv 2 (k+1) \frac{\xi \hat{V}_{\xi m^2}^{(k+1)} - \hat{V}_{m^2}^{(k+1)} - \Delta V_{\xi,m^2}^{(k+1)}}{m^2} \\
&= \frac{\xi V_{\xi m^2}^{(k)} - V_{m^2}^{(k)} - 2 (k+1) \Delta V_{\xi,m^2}^{(k+1)}}{m^2} \eqend{,}
\end{split} \\
\delta W_{\xi, m^2}^{(k+1)} &\equiv 2 (k+1) \frac{\xi \hat{W}_{\xi m^2}^{(k+1)} - \hat{W}_{m^2}^{(k+1)} - \Delta W_{\xi,m^2}^{(k+1)}}{m^2} \eqend{.}
\end{align}\end{subequations}

Finally, the $W$ coefficients of the Hadamard expansion of the graviton in a general gauge read
\begin{widetext}
\begin{equation}\begin{split}
W^{\xi,\zeta(k)}_{\mu\nu\alpha'\beta'} &= W^{1,1(k)}_{\mu\nu\alpha'\beta'} - 2 (\xi-1) \left[ \nabla_{\alpha'} \nabla_{(\mu} \hat{W}^{\mathfrak{m}^2,1(k)}_{\nu)\beta'} + \nabla_{\beta'} \nabla_{(\mu} \hat{W}^{\mathfrak{m}^2,1(k)}_{\nu)\alpha'} \right] \\
&\quad- 2 (\xi-1) (k+1) \left[ (\sigma\sdot\nabla)_{\alpha'(\mu} \hat{W}^{\mathfrak{m}^2,1(k+1)}_{\nu)\beta'} + (\sigma\sdot\nabla)_{\beta'(\mu} \hat{W}^{\mathfrak{m}^2,1(k+1)}_{\nu)\alpha'} + 2 (k+2) \sigma_{(\mu} \hat{W}^{\mathfrak{m}^2,1(k+2)}_{\nu)(\alpha'} \sigma_{\beta')} \right] \\
&\quad+ \frac{2 (1-\zeta)}{(1-2\zeta)} \left[ (k+2) (k+1) (g\sdot\sigma^2)_{\mu\nu\alpha'\beta'} \frac{W^{(k+2)}_{\mathbf{M}^2} - W^{(k+2)}_{\mathfrak{m}^2}}{\mathbf{M}^2 - \mathfrak{m}^2} + (k+1) (g\sdot\sigma\sdot\nabla)_{\mu\nu\alpha'\beta'} \frac{W^{(k+1)}_{\mathbf{M}^2} - W^{(k+1)}_{\mathfrak{m}^2}}{\mathbf{M}^2 - \mathfrak{m}^2} \right] \\
&\quad+ \frac{2 (1-\zeta)}{(1-2\zeta)} \left[ \left( g_{\alpha'\beta'} \nabla_\mu \nabla_\nu + g_{\mu\nu} \nabla_{\alpha'} \nabla_{\beta'} \right) \frac{W^{(k)}_{\mathbf{M}^2} - W^{(k)}_{\mathfrak{m}^2}}{\mathbf{M}^2 - \mathfrak{m}^2} \right] \\
&\quad+ \nabla_\mu \nabla_\nu \nabla_{\alpha'} \nabla_{\beta'} \Delta W^{\xi,\zeta(k)} + (k+1) (\sigma\sdot\nabla^3)_{\mu\nu\alpha'\beta'} \Delta W^{\xi,\zeta(k+1)} + (k+2) (k+1) (\sigma^2\sdot\nabla^2)_{\mu\nu\alpha'\beta'} \Delta W^{\xi,\zeta(k+2)} \\
&\quad+ (k+3) (k+2) (k+1) (\sigma^3\sdot\nabla)_{\mu\nu\alpha'\beta'} \Delta W^{\xi,\zeta(k+3)} + (k+4) (k+3) (k+2) (k+1) \sigma_\mu \sigma_\nu \sigma_{\alpha'} \sigma_{\beta'} \Delta W^{\xi,\zeta(k+4)} \\
&\quad+ \frac{2 (1-\zeta)}{(1-2\zeta)} \left[ (2k+3) (g\sdot\sigma^2)_{\mu\nu\alpha'\beta'} \frac{V^{(k+2)}_{\mathbf{M}^2} - V^{(k+2)}_{\mathfrak{m}^2}}{\mathbf{M}^2 - \mathfrak{m}^2} + (g\sdot\sigma\sdot\nabla)_{\mu\nu\alpha'\beta'} \frac{V^{(k+1)}_{\mathbf{M}^2} - V^{(k+1)}_{\mathfrak{m}^2}}{\mathbf{M}^2 - \mathfrak{m}^2} \right] \\
&\quad- \frac{(\xi-1)}{(k+1)} \left[ \sigma_{\alpha'(\mu} V^{\mathfrak{m}^2,1(k)}_{\nu)\beta'} + \sigma_{\beta'(\mu} V^{\mathfrak{m}^2,1(k)}_{\nu)\alpha'} + \sigma_\mu \nabla_{(\alpha'} V^{\mathfrak{m}^2,1(k)}_{|\nu|\beta')} + \sigma_\nu \nabla_{(\alpha'} V^{\mathfrak{m}^2,1(k)}_{|\mu|\beta')} + 2 \nabla_{(\mu} V^{\mathfrak{m}^2,1(k)}_{\nu)(\alpha'} \sigma_{\beta')} \right] \\
&\quad- 2 (\xi-1) \frac{(2k+3)}{(k+2)} \sigma_{(\mu} V^{\mathfrak{m}^2,1(k+1)}_{\nu)(\alpha'} \sigma_{\beta')} + (\sigma\sdot\nabla^3)_{\mu\nu\alpha'\beta'} \Delta V^{\xi,\zeta(k+1)} + (2k+3) (\sigma^2\sdot\nabla^2)_{\mu\nu\alpha'\beta'} \Delta V^{\xi,\zeta(k+2)} \\
&\quad+ (3k^2+12k+11) (\sigma^3\sdot\nabla)_{\mu\nu\alpha'\beta'} \Delta V^{\xi,\zeta(k+3)} + 2 (2k+5) (k^2+5k+5) \sigma_\mu \sigma_\nu \sigma_{\alpha'} \sigma_{\beta'} \Delta V^{\xi,\zeta(k+4)}
\end{split}\end{equation}
with
\begin{equation}
\Delta W^{\xi,\zeta(k)} \equiv - (\xi-1) \frac{4 (1-\zeta)}{(1-2\zeta)} \frac{\hat{W}^{(k)}_{\mathbf{M}^2} - \hat{W}^{(k)}_{\mathfrak{m}^2}}{\mathbf{M}^2 - \mathfrak{m}^2} - \frac{4 (1-\zeta)^2}{(1-2\zeta)^2} (\xi-3) \frac{W^{(k)}_{\mathbf{M}^2} - W^{(k)}_{\mathfrak{m}^2} - \left( \mathbf{M}^2 - \mathfrak{m}^2 \right) \hat{W}^{(k)}_{\mathbf{M}^2}}{\left( \mathbf{M}^2 - \mathfrak{m}^2 \right)^2}
\end{equation}
and the abbreviations~\eqref{tensor_abbr}.
\end{widetext}

\bibliography{literature}

\end{document}